\documentclass[amssymb,nofootinbib,nobibnotes,aps,prd,preprintnumbers]{revtex4}

 \usepackage{here}

\usepackage[dvipdfmx]{graphicx}
\usepackage{amssymb,amsfonts,amsmath,bm,color,multirow,cases,empheq,hyperref}
\usepackage{mathrsfs}

\usepackage{enumerate}
\usepackage{ulem}
\usepackage{afterpage}

\hypersetup{%
 setpagesize=false,
 bookmarksnumbered=true,%
 bookmarksopen=true,%
 colorlinks=true,%
 linkcolor=blue,%
 citecolor=red}

\begin{document}




\title{Static equilibrium of  multi-black holes in expanding bubbles  in five dimensions}

\author{Shinya Tomizawa}
\email{tomizawa@toyota-ti.ac.jp}
\author{Ryotaku Suzuki}
\email{sryotaku@toyota-ti.ac.jp}
\affiliation{Mathematical Physics Laboratory, Toyota Technological Institute\\
Hisakata 2-12-1, Nagoya 468-8511, Japan}

\date{\today}

\preprint{TTI-MATHPHYS-26}




\begin{abstract} 

We investigate possible configurations for vacuum multi-black holes that maintain static equilibrium in expanding bubbles. 
Our analysis assumes a five-dimensional Weyl metric to describe the spacetime, facilitating the derivation of solutions based on the provided rod structure. 
We consider a spacetime having expanding bubbles  caused by one or two acceleration horizons, and show that various configurations such as two bubbles, four bubbles devoid of horizons, a black saturn, a black di-ring, a bicycling black ring (orthogonal black di-ring), and a five-dimensional black hole binary can achieve equilibrium within expanding bubbles. 
Specifically, we demonstrate that equilibrium requires two acceleration horizons on both sides for the bicycling ring and the five-dimensional black hole binary. 
However, only one acceleration horizon is necessary for achieving equilibrium in the case of the black saturn and the black di-ring.

  \end{abstract}

\date{\today}
\maketitle



\section{Introduction}

Despite recent significant development in solution-generating techniques of black hole solutions, our comprehension of higher-dimensional black holes remains limited due to its complexity and the multitude of degrees of freedom it captures, 
since in higher-dimensional spacetime, the interaction among various sectors such as gravity, rotations, electric or magnetic fields, and de Sitter expansion (Anti de Sitter contraction) enriches the dynamics and phases of black holes~\cite{Emparan:2008eg}.
In particular, asymptotically flat, stationary, bi-axisymmetric five-dimensional black holes are proven to admit horizons with spatial cross-sections of  a ring $S^2 \times S^1$ or a lens space $L(p;q)$ (where $p,q$ are coprime integers) as well as a sphere $S^3$~\cite{Galloway:2005mf,Cai:2001su,Hollands:2007aj,Hollands:2010qy}. 
In the vacuum Einstein theory, there are several solutions with corresponding topologies, each revealing unique properties of black hole configurations. 
For instance, the Myers-Perry black holes represent spacetimes with topology $S^3$~\cite{Myers:1986un}, 
and black rings exhibit a topology of $S^2 \times S^1$~\cite{Emparan:2001wn,Pomeransky:2006bd}. 
On the other hand, previous attempts to find the vacuum black lens a topology of $L(p;q)$ have been met with challenges and were ultimately unsuccessful~\cite{Evslin:2008gx,Chen:2008fa,Tomizawa:2019acu}, and its nonexistence was suggested~\cite{Lucietti:2020phh}.
However, it is noteworthy that the black lenses with an $L(n;1)$-horizon (where $n$ is an integer with $n\ge 2$) were at least discovered as a supersymmetric solution~\cite{Kunduri:2014kja,Tomizawa:2016kjh,Breunholder:2017ubu}, highlighting the potential existence of such black hole configurations within a class of non-supersymmetric solutions. 
Additionally, there are multi-black hole solutions describing the black saturn, which is a superposition of $S^3$ and $S^2 \times S^1$, and the black di-ring or bicycling black ring (orthogonal black di-ring), which is a superposition of two $S^2 \times S^1$~\cite{Elvang:2007rd,Iguchi:2007is,Izumi:2007qx,Elvang:2007hs}.

\medskip

Exact solutions of the Einstein equations describing multiple black holes are of significant interest in physics, since they allow us to gain insights into the interactions between these objects. 
However, due to a lack of symmetry, solving the Einstein equations and deriving such solutions poses a challenging problem. 
Despite this difficulty, there are several previous studies on the topic.
As a first example, Israel and Khan~\cite{Israel1964}  found the exact solution that describes an arbitrary number of Schwarzschild black holes arranged along a rotational  axis. 
These black holes remain in static equilibrium thanks to conical singularities extending between them, which are crucial for counterbalancing their gravitational attraction.
Tan and Teo~\cite{Tan:2003jz}  used the generalized Weyl formalism to construct an asymptotically flat static vacuum Einstein solution, representing a superposition of multiple five-dimensional Schwarzschild black holes as a five-dimensional analog of the Israel-Khan solution~\cite{Israel1964}. 
The equilibrium of these black holes is still maintained only by conical singularities along the two rotational axes.
Thus, vacuum static  solutions are plagued by unavoidable conical singularities because, in a vacuum case, there exists no force capable of counterbalancing gravity.
This fact has been demonstrated through analyses of static, asymptotically flat, vacuum black hole solutions in Einstein theory in four or more dimensions, revealing the uniqueness of the Schwarzschild solution and hence, the absence of multi-black hole solutions~\cite{Israel:1967wq,Bunting1987,Gibbons:2002bh}. 
This leads to the intriguing question of whether the centrifugal (repulsive) force generated by rotation can effectively balance the gravitational attraction.
In a related development, Herdeiro et al.~\cite{Herdeiro:2008en} utilized the inverse scattering method to create an exact solution depicting the superposition of two Myers-Perry black holes, analogous to the double Kerr solution~\cite{Kramer1980} in five dimensions.
Each black hole in this solution has a single angular momentum parameter, and they are aligned in the same plane. 
However, this solution generally also exhibits unavoidable conical singularities in both spatial 2-planes.

\medskip
However, when including charges, static equilibrium of multi-black hole  is allowed. 
For instance, the Majumdar-Papapetrou solution~\cite{Majumdar:1947eu,Papapetrou1947} is a well-known exact solution in Einstein-Maxwell theory which describes a configuration of multiple charged black holes in static equilibrium.
In this solution, the gravitational attraction between the charged black holes is exactly balanced by the repulsion due to their electric charges. 
As a result, these black holes can remain static and in equilibrium without collapsing into a single massive black hole or flying apart due to the electromagnetic repulsion. 
Thus, the Majumdar-Papapetrou solution plays an important theoretical role in general relativity 
by demonstrating how the interplay between gravity and electromagnetism can lead to stable configurations of multiple black holes or charged objects in equilibrium.
This was also generalized to asymptotically flat solutions  in higher-dimensional Einstein-Maxwell theory~\cite{Myers:1986rx} and asymptotically Kaluza-Klein solutions with a compact dimension in five-dimensional Einstein-Maxwell theory~\cite{Ishihara:2006iv}. 
As for non-spherical black holes in five dimensions, the interplay between centrifugal force from rotation and gravitational attraction can permit the existence of multi-black hole configurations like the black saturn, black di-ring and bicycling black ring~\cite{Iguchi:2007is,Izumi:2007qx,Elvang:2007hs}.
This highlights the significant role of electric charge and rotation in facilitating the presence of multi-black hole systems.
Moreover, a recent development introduced the first multi-black hole solution within four-dimensional Einstein theory featuring a positive cosmological constant, without the presence of charge or rotation~\cite{Dias:2023rde}. 
 In this case, static equilibrium is achieved by the force balance of the gravitational attraction between two black holes and cosmic expansion, marking a notable instance of non-uniqueness in four-dimensional Einstein theory without charge and rotation.

\medskip
A notable contrast between five-dimensional and four-dimensional spacetime lies in the presence of horizonless structures known as ``bubble of nothing" or simply ``bubble" ~\cite{Witten:1981gj}, which also act as gravitational solitons in higher-dimensional gravity theories. 
In five-dimensional Kalzua-Klein theory, the regular exact solutions that describe two black holes in static equilibrium on bubbles were constructed~\cite{Elvang:2002br,Tomizawa:2007mz,Iguchi:2007xs}.
Recent studies have revealed that an expanding bubble connected to an acceleration horizon can exert an expansion force on the enclosed spacetime, maintaining static equilibrium in various non-rotating setups like five-dimensional non-rotating black rings and four-dimensional non-rotating black hole binaries~\cite{Astorino:2022fge}.
This expansion force can be roughly likened to that experienced in de Sitter spacetime, suggesting that such configurations could potentially withstand de Sitter expansion. 
The generalized Weyl ansatz offers a straightforward approach to obtaining solutions for black holes within expanding bubbles in equilibrium~\cite{Emparan:2001wk}, facilitating the search for de Sitter solutions in similar configurations.
The formations of expanding bubbles in these dimensions emerge as a result of the limiting process of static black hole binaries and black rings, respectively. 
This limit is characterized by a scenario where the separation between the two black holes or the inner hole of the black ring diminishes significantly, leading to the acceleration horizons of the bubbles aligning with the horizons of the black holes. Furthermore, they demonstrated that bubble spacetimes have the capacity to accommodate black hole binaries and black rings in static equilibrium, where their gravitational attraction is counteracted by the expansion of the background spacetime.
However, previous research has demonstrated that the expansion force generated by expanding bubbles is insufficient to balance non-rotating black lenses~\cite{Tomizawa:2022qyd}.

\medskip
In this study, our attention is directed towards various five-dimensional multi-black hole configurations without matter, rotation and cosmological constant,  that are balanced in expanding bubbles, extending the analysis from Ref.~\cite{Astorino:2022fge}. 
Specifically, we investigate whether a black saturn, a  black di-ring, a bicycling black ring, and a black hole binary can achieve equilibrium inside expanding bubbles without the need for rotation and charge, although rotation is crucial for the first three, and charge for the fourth in an asymptotically flat case.
Additionally, we emphasize that several bubbles of nothing, without black holes, can also maintain static equilibrium within expanding bubbles.
In this paper, we work with the assumption that the spacetime is described by the five-dimensional Weyl metric~\cite{Emparan:2001wk}, which allows us to derive exact solutions based on the provided rod structure.
We focus on static bi-axisymmetric spacetime configurations with one or two acceleration horizons attached to either one or both ends of the rod structure.
Our study reveals that these multi-black holes
 can achieve static equilibrium
 in expanding bubbles by balancing the gravitational force between two black hole horizons and  the pull force by
the acceleration horizon. 
In particular, we demonstrate that static equilibrium requires two acceleration horizons on both sides for the bicycling black ring and the five-dimensional black hole binary, but it requires only one acceleration horizon for the black saturn and the  black di-ring.

\medskip
The remainder of this paper is organized as follows: 
Firstly, in Sec.~\ref{sec:setup}, we introduce the foundational framework and provide a concise overview of the generalized Weyl solution in five dimensions. This solution pertains to static and axisymmetric vacuum spacetime characterized by three commuting Killing vector fields.
Subsequently, in Sec.~\ref{sec:bubble}, we consider static equilibrium of  expanding bubbles of nothing without horizons. 
In Sec.~\ref{sec:black holes}, we delve into various configurations involving multiple black holes that are balanced within expanding bubbles. The metric can be easily read off from a so-called ``rod diagram"\cite{Emparan:2001wk,Harmark:2004rm}.
This section delves into the conditions necessary and sufficient for achieving static equilibrium of various multi-black holes within expanding bubbles in expanding bubbles,  a black saturn in Sec.~\ref{sec:bs}, a  black di-ring in Sec.~\ref{sec:bdr}, a bicycling black ring in Sec.~\ref{sec:obdr}, and a five-dimensional black hole binary in Sec.~\ref{sec:5dbbh}. 
Lastly, we provide a summary of our findings and conclusions drawn from the study in Sec.~\ref{sec:summary}.

\section{Setup}\label{sec:setup}
In this paper, we investigate the multi-black hole solutions of the five-dimensional vacuum Einstein theory which are in static equilibrium inside expanding bubbles of nothing. 
Specifically, we examine a class of solutions known as the generalized Weyl solution~\cite{Emparan:2001wk}.
Let us consider a five-dimensional static and axisymmetric vacuum spacetime admitting three commuting Killing vector fields $V_{(i)}$ $(i = 1, 2, 3)$, following the argument in previous works~\cite{Emparan:2001wk}. 
The commutativity of the Killing vectors, denoted by $[V_{(i)}, V_{(j)}] = 0$, allows us to establish a coordinate system such that $V_{(i)} = \partial/\partial x^i$ $(i = 1, 2, 3)$, with the metric being independent of the coordinates $x^i$. 
Here, $\partial/\partial x^1$ represents the timelike Killing vector, while $\partial/\partial x^2$ and $\partial/\partial x^3$ denote the spacelike rotational Killing vectors, and we denote the coordinate $x^i$ as $(x^1,x^2,x^3)=(t,\phi,\psi)$.

\medskip
With these assumptions, the two-dimensional space orthogonal to all three Killing vectors becomes integrable, enabling us to express the metric as
\begin{eqnarray}
ds^2=g_{ij}(\rho,z)dx^idx^j+f(\rho,z)(d\rho^2+dz^2),
\end{eqnarray}
together with the constraint 
\begin{eqnarray}
{\rm det} (g_{ij})=-\rho^2,
\end{eqnarray}
where the three-dimensional diagonal metric $g_{ij}= {\rm diag}(g_0,g_1,g_2)$ and the function $f$ do not depend on $x^i$. 
Then, it follows from the Einstein equation $R_{ij}=0$ that the metric $g_{ij}$ reduces to the harmonic equations on a three-dimensional Euclid space in the abstract cylindrical coordinate system, defined as $\gamma:= d\rho^2 +dz^2+ \rho^2 d\varphi ^2$,
\begin{align}
\triangle_\gamma \log g_k  =0,\quad k=0,1,2,\label{eq:weyl-harmonic}
\end{align}
and moreover, from the other components of the Einstein equation $R_{\rho\rho}-R_{zz}=R_{\rho z}=0$, the function $f$ is determined up to a constant by 
\begin{align}
\partial_\rho \log f =-\frac{1}{\rho} +  \frac{1}{4\rho} \sum_{k=0}^2 g_k^{-2} [ (\partial_\rho g_k)^2-(\partial_z g_k)^2],\quad \partial_z \log f = \frac{1}{2\rho}\sum_{k=0}^2 g_k^{-2} \partial_\rho g_k \partial_z g_k,  \label{eq:f}
\end{align}
where the integrability condition $\partial_\rho \partial_z  f = \partial_z \partial_\rho  f$ is derived from Eq.~(\ref{eq:weyl-harmonic}).
The solution for Eq.~(\ref{eq:weyl-harmonic}) is easily constructed in terms of $\rho^2$, $\mu_i$ and $\bar{\mu}_i$, where
\begin{eqnarray}
\mu_i&=&\sqrt{\rho^2+(z-z_i)^2}-(z-z_i),\quad \bar{\mu}_i=-\sqrt{\rho^2+(z-z_i)^2}-(z-z_i).
\end{eqnarray}
The function $\log \mu_i$ is one of the solutions to the harmonic equation~(\ref{eq:weyl-harmonic}) for the semi-infinite rod $z \in [z_i,\infty)$ with mass density $1/2$ on the $z$-axis and $\log \bar{\mu}_i$ for the semi-infinite rod $z \in (-\infty , z_i]$. 
More general solutions can be obtained by taking the product of $\rho^2$, $\mu_i$  and $\bar{\mu}_i$,
which is determined by the mass distribution along the $z$-axis, known as the ``rod structure''~\cite{Emparan:2001wk,Harmark:2004rm}. 
Once $g_{ij}$ is determined, the solution $f$ to Eq.~(\ref{eq:f}) can be obtained, following the procedure outlined in the appendix of Ref.~\cite{Izumi:2007qx}.

\medskip

Additionally, a ``rod diagram" is a graphical representation of the rod structure, used often in the study of the construction of higher-dimensional black hole solutions, particularly in the context of the Weyl metric~\cite{Emparan:2001wk,Harmark:2004rm}. 
It is a visualization technique that simplifies the description of the spacetime geometry near black holes, 
focusing on the essential features such as the presence of black holes, acceleration horizons, expanding bubbles and rotational axes.
In a rod diagram, each black hole is represented by a finite line segment for $g_0$ called a timelike rod, with the length of the rod corresponding to the mass or charge of the black hole. 
Acceleration horizons, which are associated with the presence of expanding bubbles, are represented by additional semi-infinite line segments attached to the timelike rods. 
Furthermore, expanding bubbles and rotational axes are  represented by line segments for $g_1$ or $g_2$  called spacelike rods.
In summary, a rod diagram provides us a clear and intuitive way to visualize the complex geometry of spacetimes with multiple black holes and expanding bubbles, making it a valuable tool in theoretical studies of gravitational physics.

\section{Bubbles in expanding bubbles}\label{sec:bubble}

In this section, we consider two horizonless spacetimes with only bubbles of nothing.

\subsection{Expanding bubbles of nothing}

\begin{figure}[H]

\begin{minipage}[t]{0.5\columnwidth}
    \centering
 \includegraphics[width=8.5cm]{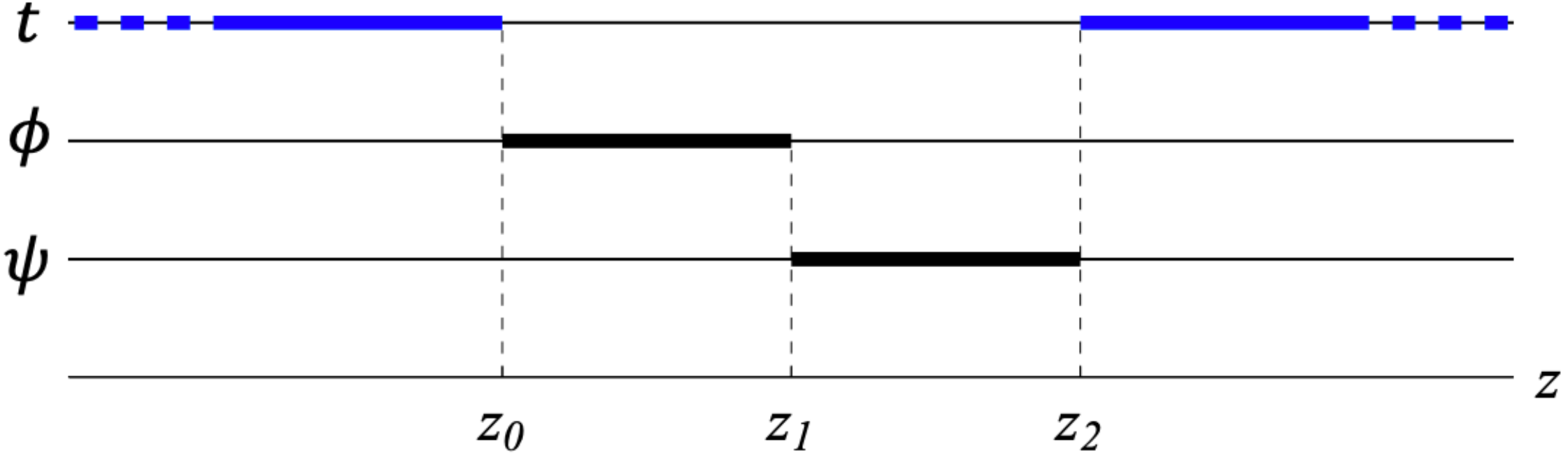}
  \caption{Rod diagram of 5D expanding bubble of nothing. \label{fig:rod-A} }
  \end{minipage}
   \begin{minipage}[t]{0.5\columnwidth}
    \centering
 \includegraphics[width=8.5cm]{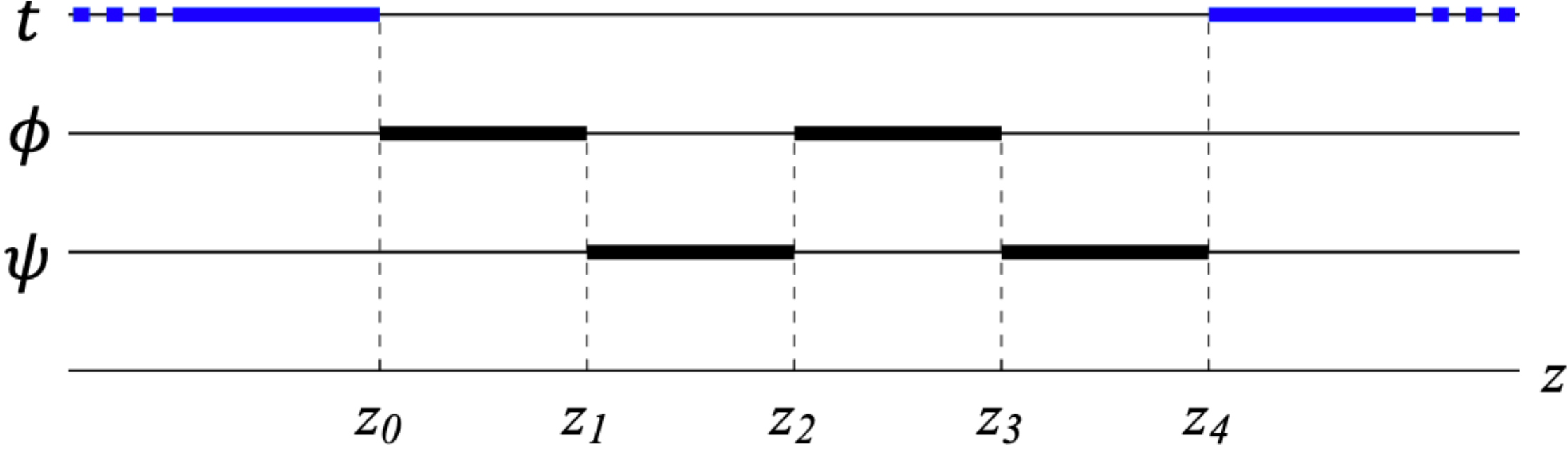}
  \caption{Rod diagram  of two bubbles inside 5D expanding bubble of nothing. \label{fig:rod-B} }
  \end{minipage}

\end{figure}

First, as a simplest example of only expanding bubbles,  let us consider the spacetime with the rod structure given by Fig.~\ref{fig:rod-A}, from which the metric can be read off as
\begin{eqnarray}
ds^2=-\frac{\rho^2\mu_2}{\mu_0}dt^2
+\frac{\mu_0}{\mu_1}d\phi^2
+\frac{\mu_1}{\mu_2}d\psi^2+
f(d\rho^2+dz^2),
\end{eqnarray}
with
\begin{eqnarray}
f=C_f \frac{\mu_2}{\mu_0}\frac{R_{01} R_{02} R_{12} }
{R_{00}R_{11}R_{22}  }
\end{eqnarray}
where $R_{ij}:=\rho^2+\mu_i\mu_j$ and $C_f$ is an arbitrary constant. 
This solution was first studied in Ref.~\cite{Emparan:2001wk} and then
 can be obtained in Ref.~\cite{Astorino:2022fge} by the limit at which the two black holes become infinitely large  for the five-dimensional asymptotically flat, static black hole binary with inevitable conical singularities~\cite{Tan:2003jz}. 
This spacetime admits the presence of two accelerating horizons, which correspond to the timelike semi-inifinite rods $(-\infty,z_0]$ and $[z_2,\infty)$. 
The two spacelike rods $[z_0,z_1]$ and $[z_1,z_2]$ represent bubbles, each having topology $D^2$, which are  orthogonal and intersect at the point $z=z_1$. Meanwhile, the timelike semi-inifinite rods $(-\infty,z_0]$ and $[z_2,\infty)$ represent the bubble acceleration horizons with topology $S^1\times {\mathbb R}^2$ .

\medskip
Since in general, the spacetime includes conical singularities in two rods $[z_0,z_1]$ and $[z_1,z_2]$ on the $z$-axis,  we eliminate them by imposing the regularity conditions as follows: 

\begin{itemize}

\item[(i)] On the spacelike finite rod  $[z_0,z_1]$, which represents an expanding bubble corresponding to  $\phi$-rotational axis,
the absence of conical singularities requires the periodicity of $\phi$, 
\begin{eqnarray}
\lim_{\rho\to 0}\sqrt{\frac{\rho^2 f}{g_{\phi \phi } }}=\frac{\Delta\phi}{2\pi} \Longleftrightarrow 4C_fz_{10}z_{20}=\left( \frac{\Delta\phi}{2\pi}\right)^2. \label{eq:bubble1_01}
\end{eqnarray}

\item[(ii)] On the spacelike finite rod $[z_1,z_2]$, which represents the bubble corresponding to $\psi$-rotational axis, similarly, the absence of conical singularities requires the periodicity of $\psi$,
\begin{eqnarray}
\lim_{\rho\to 0}\sqrt{\frac{\rho^2 f}{g_{\psi \psi } }}=\frac{\Delta\psi}{2\pi} \Longleftrightarrow 4C_fz_{20}z_{21}=\left( \frac{\Delta\psi}{2\pi}\right)^2 .\label{eq:bubble1_12}
\end{eqnarray}
\end{itemize}
In order to solve these, we choose the parametrization of the rod endpoints as
\begin{eqnarray}
l_1:=z_{21},\quad l_2:=z_{10},\quad C_f=\frac{1}{4},
\end{eqnarray}
where $l_1$ and $l_2$ present the size of the bubbles $[z_1,z_2]$ and $[z_0,z_1]$, respectively. 
Then Eqs.~(\ref{eq:bubble1_01}) and (\ref{eq:bubble1_12}) simply set the periodicity of $\psi$ and $\phi$, respectively, as
\begin{eqnarray}
l_1(l_1+l_2)=\left( \frac{\Delta\psi}{2\pi}\right)^2,\quad 
l_2(l_1+l_2)=\left( \frac{\Delta\phi}{2\pi}\right)^2, \label{eq:bubble}
\end{eqnarray}
without further constraints.
Since Eq.~(\ref{eq:bubble}) simply determines the periodicities of $\phi$ and $\psi$,  both the bubble sizes $l_1$ and $l_2$ can have any values,  with conical singularities on both the bubbles being absent,  and  hence two bubble can have arbitrary size.
This spacetime reduces to the Minkowski spacetime at the limit $l_1,l_2\to\infty$ with rescaling $\phi \to \sqrt{l_2(l_1+l_2)} \phi,\ \psi \to \sqrt{l_1(l_1+l_2)} \psi$, while keeping the absence of conical singularities.

\subsection{Bubbles in expanding bubbles}

Next, as a second example, we consider ``two bubbles inside two expanding bubbles of nothing" whose rod diagram is given by the right panel in Fig.~\ref{fig:rod-A}. 
From this rod diagram, we can construct the metric as
\begin{eqnarray}
ds^2=-\frac{\rho^2\mu_4}{\mu_0}dt^2
+\frac{\mu_0\mu_2}{\mu_1\mu_3}d\phi^2
+\frac{\mu_1\mu_3}{\mu_2\mu_4}d\psi^2+
f(d\rho^2+dz^2),
\end{eqnarray}
with 
\begin{eqnarray}
f=C_f \frac{\mu_4}{\mu_0}\frac{R_{01}R_{03}R_{04}  R_{12}^2 R_{14} R_{23}^2 R_{34} }
{R_{00}R_{02}  R_{11}R_{13}^2 R_{22}R_{24}R_{33} R_{44}    }.
\end{eqnarray}
This spacetime also admits the presence of two bubble acceleration horizons with topology $S^1\times {\mathbb R}^2$ on the timelike semi-inifinite rods $(-\infty,z_0]$ and $[z_4,\infty)$, and two expanding bubbles on the spacelike finite rods $[z_0,z_1]$, $[z_3,z_4]$ with topology $D^2$, and two additional bubbles on the spacelike finite rods $[z_1,z_2]$, $[z_2,z_3]$ with topology $S^2$, where the latter two are orthogonal and intersect at the point $z=z_2$.

\medskip
In order to eliminate conical singularities on two rods $[z_1,z_2]$, $[z_2,z_3]$  on the $z$-axis as well as $[z_0,z_1]$, $[z_3,z_4]$,  we impose the regularity conditions as follows: 
\begin{itemize}

\item[(i)] On the spacelike finite rod $[z_0,z_1]$ which  represents an expanding bubble corresponding to  $\phi$-rotational axis, the absence of conical singularities requires the periodicity of $\phi$, 
\begin{eqnarray}
\lim_{\rho\to 0}\sqrt{\frac{\rho^2 f}{g_{\phi \phi } }}=\frac{\Delta\phi}{2\pi} \Longleftrightarrow 4C_f \frac{z_{10}z_{30}z_{40}}{z_{20}}=\left( \frac{\Delta\phi}{2\pi}\right)^2. \label{eq:bubble2_01}
\end{eqnarray}

\item[(ii)]  On the spacelike finite rod $[z_1,z_2]$ which  represents a bubble corresponding to  $\psi$-rotational axis, the absence of conical singularities requires the periodicity of
\begin{eqnarray}
\lim_{\rho\to 0}\sqrt{\frac{\rho^2 f}{g_{\psi \psi } }}=\frac{\Delta\psi}{2\pi} \Longleftrightarrow 4C_f \frac{z_{21}^2z_{30}z_{40}z_{41}} {z_{20}z_{31}^2}=\left( \frac{\Delta\psi}{2\pi}\right)^2, \label{eq:bubble2_12}
\end{eqnarray}

\item[(iii)]  On the spacelike finite rod $[z_2,z_3]$ which  represents a bubble corresponding to  $\phi$-rotational axis, the absence of conical singularities requires the periodicity of $\phi$, 
\begin{eqnarray}
\lim_{\rho\to 0}\sqrt{\frac{\rho^2 f}{g_{\phi \phi } }}=\frac{\Delta\phi}{2\pi} \Longleftrightarrow 4C_f \frac{z_{30}z_{32}^2z_{40}z_{41}} {z_{31}^2z_{42}}=\left( \frac{\Delta\phi}{2\pi}\right)^2, \label{eq:bubble2_23}
\end{eqnarray}

\item[(iv)]  On the spacelike finite rod $[z_3,z_4]$ which  represents an expanding bubble corresponding to  $\psi$-rotational axis, the absence of conical singularities requires the periodicity of
\begin{eqnarray}
\lim_{\rho\to 0}\sqrt{\frac{\rho^2 f}{g_{\psi \psi } }}=\frac{\Delta\psi}{2\pi} \Longleftrightarrow 4C_f \frac{z_{40}z_{41}z_{43}} {z_{42}}=\left( \frac{\Delta\psi}{2\pi}\right)^2,\label{eq:bubble2_34}
\end{eqnarray}

\end{itemize}
To avoid conical singularities on $\phi$ and $\psi$ rotational axes simultaneously,
from Eqs.~(\ref{eq:bubble2_01}),  (\ref{eq:bubble2_23}) and  Eqs.~(\ref{eq:bubble2_12}), (\ref{eq:bubble2_34}), we have 
\begin{align}
\left( \frac{\Delta\phi}{2\pi}\right)^2  =   \frac{z_{10}z_{30}z_{40}}{z_{20}}=
  \frac{z_{30}z_{32}^2z_{40}z_{41}} {z_{31}^2z_{42}}, \label{eq:bubble2-conifree1}
\end{align}
and 
\begin{align}
\left( \frac{\Delta\psi}{2\pi}\right)^2=  \frac{z_{40}z_{41}z_{43}} {z_{42}}=\frac{z_{21}^2z_{30}z_{40}z_{41}} {z_{20}z_{31}^2},
\label{eq:bubble2-conifree2}
\end{align}
where we choose $C_f=1/4$.

\medskip
Let us choose the parametrization of the rod endpoints $z_i\ (i=0,\ldots,4)$ as
\begin{align}
z_0=-(1+\mu)\ell,\quad z_1 = - \ell,\quad z_2=0, \quad z_3 = \nu \ell,\quad z_4 = (\nu+\gamma)\ell,
\end{align}
where $\mu\ell, \ell,\nu\ell$, and $\gamma \ell$ corresponds to the sizes of the bubbles on $[z_0,z_1]$,  $[z_1,z_2]$, $[z_2,z_3]$ and $[z_3,z_4]$, respectively.
Eqs.~(\ref{eq:bubble2-conifree1}) and (\ref{eq:bubble2-conifree2}) can be rewritten as, respectively, 
\begin{eqnarray}
\left( \frac{\Delta\phi}{2\pi}\right)^2 = \frac{\ell^2 \mu  (\mu +\nu +1) (\gamma +\mu +\nu +1)}{\mu +1}=\frac{\ell^2 \nu ^2
   (\gamma +\nu +1) (\mu +\nu +1) (\gamma +\mu +\nu +1)}{(\nu +1)^2 (\gamma +\nu
   )},\label{eq:bubble2-conifree1b}
\end{eqnarray}
\begin{eqnarray}
\left( \frac{\Delta\psi}{2\pi}\right)^2=\frac{\ell^2 (\gamma +\nu +1) (\mu +\nu +1) (\gamma +\mu +\nu +1)}{(\mu +1) (\nu
   +1)^2}=\frac{\gamma  \ell^2 (\gamma +\nu +1) (\gamma +\mu +\nu +1)}{\gamma +\nu   }.
   \label{eq:bubble2-conifree2b}
\end{eqnarray}
From the above equations, we obtain the relation on the size of bubbles,  
\begin{align}
\mu = \frac{\nu ^2 (\nu +1) (\nu +2)}{\nu ^3+3 \nu ^2+4 \nu +1},\quad \gamma =
   \frac{(\nu +1) (2 \nu +1)}{\nu ^3+4 \nu ^2+3 \nu +1},\label{eq:bubble2-sol}
\end{align}
from which the sizes of the bubble on $[z_0,z_1]$ and the  expanding bubble on $[z_3,z_4]$ are determined by the size of  the bubble on $[z_2,z_3]$.
Substituting this into Eqs.~(\ref{eq:bubble2-conifree1b}) and (\ref{eq:bubble2-conifree2b}), we obtain
\begin{align}
 &\left( \frac{\Delta\phi}{2\pi}\right)^2=\frac{\ell^2 \nu ^2 (\nu +1)^5 (\nu +2)^2 (2 \nu +1)^2}{\left(\nu ^3+3 \nu ^2+4 \nu +1\right)^2 \left(\nu
   ^3+4 \nu ^2+3 \nu +1\right)},\\
& \left( \frac{\Delta\psi}{2\pi}\right)^2=\frac{\ell^2 (\nu +1)^5 (\nu +2)^2 (2 \nu +1)^2}{\left(\nu ^3+3 \nu ^2+4 \nu +1\right) \left(\nu ^3+4 \nu
   ^2+3 \nu +1\right)^2}, 
\end{align}
which determine the periodicities of $\phi$ and $\psi$.

\medskip

Therefore, we can observe that the presence of two bubbles between two expanding bubbles  requires the existence of both acceleration horizons on $(-\infty,z_0]$ and $[z_4,\infty)$.
One can observe that the presence of both acceleration horizons on $(-\infty,z_0]$ and $[z_4,\infty)$ is necessary for balance, as $\mu$ and $\gamma$ given by Eq.~(\ref{eq:bubble2-sol}) remain finite for any finite $\nu>0$. More precisely, one cannot take the limit $\gamma \to \infty$ or $\mu \to \infty$ while keeping $\nu$ finite, as this corresponds to the existence of a single acceleration horizon. Therefore, we can conclude that the presence of two bubbles between two expanding bubbles requires the existence of both acceleration horizons on $(-\infty,z_0]$ and $[z_4,\infty)$.

\section{Multi-black holes in expanding bubbles}\label{sec:black holes}
Subsequently, in this section, we consider four configurations of multiple black holes that are balanced within expanding bubbles. 
The discussions focus on a black saturn in Sec.~\ref{sec:bs}, a black di-ring in Sec.~\ref{sec:bdr}, a bicycling black ring in Sec.~\ref{sec:obdr}, and a five-dimensional black hole binary in Sec.~ \ref{sec:5dbbh}.

\subsection{Black saturn in expanding bubbles}\label{sec:bs}
\begin{figure}[H]
 \centering
\includegraphics[width=9cm]{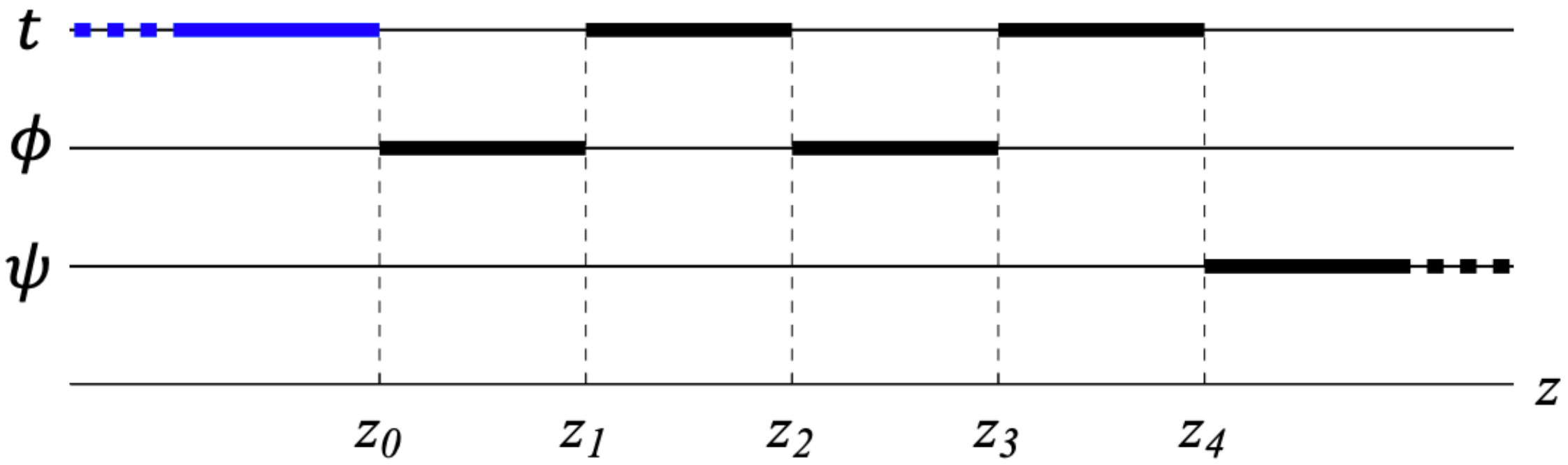}
\caption{Rod structure of a black saturn inside bubble of nothing. }
\label{fig:bs-rod}
\end{figure}

First, let us consider ``a static black saturn in expanding bubbles", for which the rod diagram is depicted in Fig.~\ref{fig:bs-rod}.
Based on this rod diagram, we can construct the metric as follows:

\begin{eqnarray}
ds^2=-\frac{\rho^2\mu_1\mu_3}{\mu_0\mu_2\mu_4}dt^2
+\frac{\mu_0\mu_2}{\mu_1\mu_3}d\phi^2
+\mu_4d\psi^2+
f(d\rho^2+dz^2),
\end{eqnarray}
with
\begin{eqnarray}
f=C_f \frac{\mu_4R_{01}^2R_{03}^2R_{12}^2 R_{14} R_{23}^2 R_{34}}{R_{00}R_{02}^2R_{04}R_{11}R_{13}^2 R_{22}R_{24}R_{33} R_{44} }.
\end{eqnarray}

A black hole horizon with topology $S^3$ exists on the timelike finite rod $[z_3,z_4]$, and it is surrounded by a black ring horizon with topology $S^1\times S^2$ on the timelike finite rod $[z_1,z_2]$. Additionally, this spacetime also features a single bubble acceleration horizon with topology $S^1\times \mathbb{R}^2$ on the timelike semi-infinite rods $(-\infty,z_0]$, along with an expanding bubble and another bubble on the spacelike finite rods $[z_0,z_1]$ and $[z_2,z_3]$, which have the cylinder topology of $S^1\times \mathbb{R}$.

\medskip
In order to eliminate conical singularities on two rods $[z_0,z_1]$, $[z_2,z_3]$  on the $z$-axis, we impose the regularity conditions as, respectively,  
\begin{eqnarray}
&&\lim_{\rho\to 0}\sqrt{\frac{\rho^2 f}{g_{\phi \phi } }}=\frac{\Delta\phi}{2\pi} \Longleftrightarrow 2C_f\frac{z_{10}^2z_{30}^2 }{z_{20}^2 z_{40} }=\left( \frac{\Delta\phi}{2\pi}\right)^2,\label{eq:BS1:conical01}\\
&&\lim_{\rho\to 0}\sqrt{\frac{\rho^2 f}{g_{\phi \phi } }}=\frac{\Delta\phi}{2\pi} \Longleftrightarrow 2C_f\frac{z_{30}^2z_{32}^2 z_{41}  }{z_{31}^2 z_{40}  z_{42}}=\left( \frac{\Delta\phi}{2\pi}\right)^2. \label{eq:BS1:conical23}
\end{eqnarray}
The absence of conical singularities on $[z_4,\infty)$ requires the condition
\begin{eqnarray}
\lim_{\rho\to 0}\sqrt{\frac{\rho^2 f}{g_{\psi \psi } }}=\frac{\Delta\psi}{2\pi} \Longleftrightarrow C_f=\left( \frac{\Delta\psi}{2\pi}\right)^2, \label{eq:BS1:conical45}
\end{eqnarray}
where in the parameter choice of $C_f=1$, Eq.~(\ref{eq:BS1:conical45}) determines the periodicity of $\psi$ as
$\Delta \psi=2\pi$.
To avoid the conical singularities on both the $\phi$-rotational axes, Eqs.~(\ref{eq:BS1:conical01}) and 
(\ref{eq:BS1:conical23}) must be satisfied simultaneously 
\begin{align}
\left( \frac{\Delta\phi}{2\pi}\right)^2=\frac{2z_{30}^2z_{32}^2 z_{41}  }{z_{31}^2 z_{40}  z_{42}}
= \frac{2z_{10}^2z_{30}^2 }{z_{20}^2 z_{40} }.\label{eq:BS1:noconical-0}
\end{align}

\medskip
To give these parameters a physical interpretation, we introduce the positive parameters $(\ell,r,\mu,R)$ as follows:
\begin{eqnarray}
z_0=-\ell,\quad z_1=0,\quad z_2=r \ell,\quad z_3=(r+\mu)\ell,\quad z_4=(r+\mu+R)\ell,
\end{eqnarray}
where $\ell$ and $\mu\ell$ represent the sizes of the expanding bubble on $[z_0,z_1]$ and the bubble on $[z_2,z_3]$, respectively, while $r \ell$ and $R\ell$ correspond to the sizes of the black ring and black hole horizons, respectively.
Then in terms of these, Eq.~(\ref{eq:BS1:noconical-0}) can be expressed as
\begin{eqnarray}
\left(\frac{\Delta\phi}{2\pi}\right)^2=\frac{2\ell  (r+\mu+1)^2}{(r+1)^2 (r+\mu+R+1)}=\frac{2\ell \mu^2 (r+\mu+1)^2(r+\mu+R)}{(r+\mu)^2(\mu+R) (r+\mu+R+1)},
\end{eqnarray}
which is equivalent to
\begin{eqnarray}
R =\frac{\mu  (\mu +r) (\mu  (r+2)-1)}{(1-\mu) ((2+r) \mu+r)},\quad \left(\frac{\Delta\phi}{2\pi}\right)^2=\frac{2 \ell (1-\mu ) (\mu +r+1)^2 ((2+r) \mu+r)}{(r+1)^3 (r-(\mu -2) \mu )}, \label{eq:BS1:conical_b}
\end{eqnarray}
where $R>0$ is satisfied only in the range
\begin{eqnarray}
\frac{1}{r+2}<\mu <1.
\end{eqnarray}
The first equation in Eq.~(\ref{eq:BS1:conical_b}) determines the size of the black hole by the size of the black ring and the bubble on $[z_2,z_3]$. The second equation determines the periodicity of $\phi$.

\medskip
As is well-known, the black hole temperature $T=1/\beta$ can be derived by means of the Euclidean approach, i.e., by Wick rotation, compactifying the Euclidean time and the absence of conical singularities requires identifying the period with the inverse temperature:
\begin{eqnarray}
\lim_{\rho\to 0}\sqrt{\frac{\rho^2 f}{-g_{tt } }}=\frac{\beta}{2\pi}.
\end{eqnarray}
The temperatures $T_{\rm BR} (=:1/\beta_{\rm BR})$ and  $T_{\rm BH} (=:1/\beta_{\rm BH})$ for the black ring horizon on $[z_1,z_2]$ and the black hole horizon on  $[z_3,z_4]$ can be computed as, respectively, 
\begin{eqnarray}
\lim_{\rho\to 0}\sqrt{\frac{\rho^2 f}{-g_{tt } }}&=&\frac{\beta_{\rm BR}}{2\pi}  \Longleftrightarrow 
\frac{z_{21}^2z_{30}^2 z_{41}}{z_{20}^2z_{31}^2z_{40}}=\left(\frac{\beta_{\rm BR}}{2\pi}\right)^2,
\label{eq:BS1:conical12}\\
\lim_{\rho\to 0}\sqrt{\frac{\rho^2 f}{-g_{tt } }}&=&\frac{\beta_{\rm BH}}{2\pi}  \Longleftrightarrow \frac{z_{41}z_{43}}{z_{40}z_{42}}
=\left(\frac{\beta_{\rm BH}}{2\pi}\right)^2.\label{eq:BS1:conical34}
\end{eqnarray}
Furthermore, similarly, the temperature $T_{\rm AH} (=:1/\beta_{\rm AH})$  for the bubble acceleration horizon on $(-\infty,z_0]$  can be obtained as
\begin{eqnarray}
\lim_{\rho\to 0}\sqrt{\frac{\rho^2 f}{-g_{tt } }}=\frac{\beta_{\rm AH}}{2\pi}  \Longleftrightarrow 
1=\left(\frac{\beta_{\rm AH}}{2\pi}\right)^2.\label{eq:BS1:conical0}
\end{eqnarray}

\begin{figure}[H]
 \centering
\includegraphics[width=6cm]{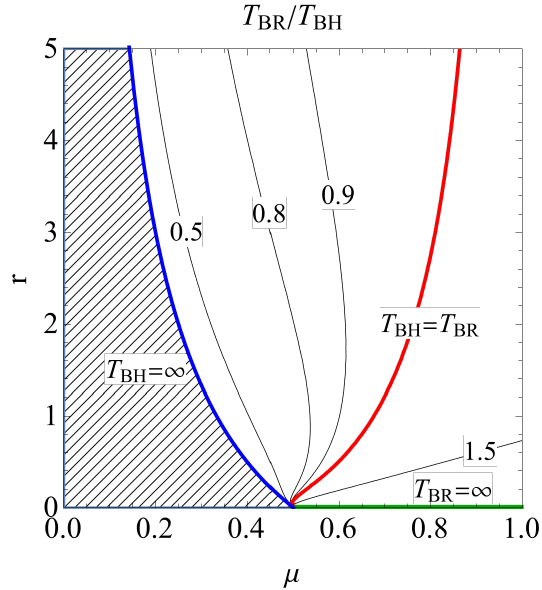}
\caption{The temperature ratio between the black ring horizon and black hole horizon for the black saturn in expanding bubble shown in the $(\mu,r)$-plane. 
The blue curve ($\mu=(r+2)^{-1}$) represents the points where $T_{\rm BH}\to\infty$ and  the black hole horizon disappears, corresponding to the static black ring in the expanding bubble. 
The green line $r=0\ (\mu>1/2)$ represents the points $T_{\rm BR}\to\infty$ and  the black ring horizon disappears, corresponding the static black hole in the expanding bubble.
There is no balanced state in the hatched region. 
The red curve corresponds to the phase where the black ring horizon and black hole horizon have the same temperature. 
\label{fig:rod}}
\end{figure}

\medskip
From Eqs.~(\ref{eq:BS1:conical34}), (\ref{eq:BS1:conical12}) and (\ref{eq:BS1:conical0}),
the inverse temperatures of each horizon can be written as
\begin{eqnarray}
\left(\frac{\beta_{\rm BH}}{2\pi}\right)^2&=&\frac{ (r+\mu)^3 ((2+r)\mu-1)}{\mu (1+r)^3 (r+(2-\mu)\mu)},\\
\left(\frac{\beta_{\rm BR}}{2\pi}\right)^2&=&  \frac{r^2(1+r+\mu)^2}{(r + 1)^3 (r+(2-\mu)\mu)},\\
   \left(\frac{\beta_{\rm AH}}{2\pi}\right)^2&=&1.
\end{eqnarray}
To discuss the phase diagram of this black saturn in  an expanding bubble of nothing, we consider the temperature ratio of the black ring horizon to that of the black hole horizon, which is given by
\begin{eqnarray}
\left(\frac{T_{\rm BR}}{T_{\rm BH}}\right)^2=\left(\frac{\beta_{\rm BH}}{\beta_{\rm BR}}\right)^2=\frac{(r+\mu)^3((2+r)\mu-1)}{r^2\mu(1+r+\mu)^2}.
\end{eqnarray}
In Fig.~\ref{fig:rod}, we illustrate the phase diagram of the static black saturn in the expanding bubble in the $(\mu,r)$-plane, accompanied by the temperature ratio profile $T_{\rm BR}/T_{\rm BH}$. According to Eq.(\ref{eq:BS1:conical_b}), as $\mu\to (r+2)^{-1}$, we have $R\to 0$ and $T_{\rm BH}\to\infty$, leading to the disappearance of the black hole horizon. Consequently, the resulting spacetime corresponds to the static black ring in an expanding bubble, as discussed in Ref.~\cite{Astorino:2022fge}. Conversely, in the $r \to 0$ limit, $T_{\rm BR}$ diverges, causing the black ring horizon to vanish, resulting in a static black hole in an expanding bubble.

One might question whether the black saturn can maintain static equilibrium even with the addition of another bubble acceleration horizon on the right end, represented as a semi-infinite timelike rod on $[z_5 , \infty)$. The answer is yes. However, the presence of only the right acceleration horizon is insufficient to support the black saturn due to unavoidable conical singularities. This implies that the gravitational attraction between the black hole and black ring cannot counterbalance the pulling force along the $\psi$-rotational axis caused by the accelerating horizon alone. Therefore, the left acceleration horizon on $(-\infty,z_0]$ is necessary for achieving static equilibrium.

\subsection{Black di-ring in expanding bubbles}\label{sec:bdr}
\begin{figure}[h]
 \centering
\includegraphics[width=9cm]{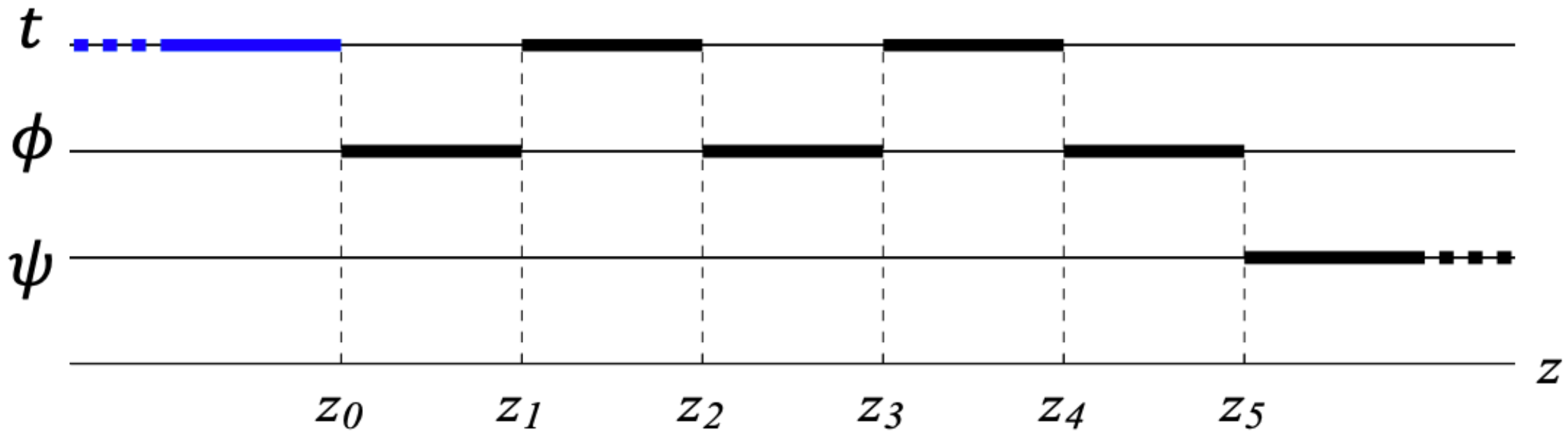}
\caption{Rod structure of a black di-ring inside bubble of nothing. }
\label{fig:rod-BDR}
\end{figure}

Next, we consider  ``a static black di-ring in expanding bubbles" for which the rod diagram is given in Fig.~\ref{fig:rod-BDR}. 
Here, from the same reason as in the black saturn, 
we assume the presence of a single acceleration bubble horizon. 
The metric can be written as
\begin{eqnarray}
ds^2=-\frac{\rho^2\mu_1\mu_3}{\mu_0\mu_2\mu_4}dt^2
+\frac{\mu_0\mu_2\mu_4}{\mu_1\mu_3\mu_5}d\phi^2
+\mu_5d\psi^2+
f(d\rho^2+dz^2),
\end{eqnarray}
with
\begin{eqnarray}
f=C_f 
\frac{\mu_5 R_{01}^2  R_{03}^2 R_{05} R_{12}^2  R_{14}^2 R_{23}^2  R_{25}   R_{34}^2   R_{45} }{R_{00}R_{02}^2R_{04}^2R_{11}R_{13}^2R_{15}R_{22}R_{24}^2R_{33}R_{35}R_{44}R_{55} }.
\end{eqnarray}

This spacetime has two black ring horizons with topology $S^1\times S^2$ on the timelike finite rods $[z_1,z_2]$ and $[z_3,z_4]$, where they are the concentric in the same plane, and the former is outside the latter.
Additionally, there is a single bubble acceleration horizon with topology $S^1\times \mathbb{R}^2$ on the timelike semi-infinite rods $(-\infty,z_0]$. Furthermore, there exists an expanding bubble with topology $S^1\times \mathbb{R}$ on the spacelike finite rods $[z_0,z_1]$, the inner $\phi$-rotational axis on $[z_2,z_3]$ with topology $S^1\times \mathbb{R}$, and the inner $\phi$-rotational axis on $[z_4,z_5]$  with topology $D^2$.

\medskip
In order to eliminate conical singularities on three rods $[z_0,z_1]$, $[z_2,z_3]$ and $[z_4,z_5]$  on the $z$-axis, we impose the regularity conditions as, respectively,  
\begin{eqnarray}
&&\lim_{\rho\to 0}\sqrt{\frac{\rho^2 f}{g_{\phi \phi } }}=\frac{\Delta\phi}{2\pi} 
\Longleftrightarrow 
2C_f
\frac{z_{10}^2z_{30}^2 z_{50} }
{z_{20}^2 z_{40}^2 }
=\left( \frac{\Delta\phi}{2\pi}\right)^2,\label{eq:BR1:conical01}\\
&&\lim_{\rho\to 0}\sqrt{\frac{\rho^2 f}{g_{\phi \phi } }}=\frac{\Delta\phi}{2\pi} 
\Longleftrightarrow 
2C_f\frac{z_{30}^2z_{32}^2z_{41}^2z_{50}z_{52}}
{z_{31}^2 z_{40}^2  z_{42}^2 z_{51} }
=\left( \frac{\Delta\phi}{2\pi}\right)^2,\label{eq:BR1:conical23}\\
&&\lim_{\rho\to 0}\sqrt{\frac{\rho^2 f}{g_{\phi \phi } }}=\frac{\Delta\phi}{2\pi} 
\Longleftrightarrow 
2C_f\frac{ z_{50}z_{52}z_{54}  }
{z_{51}z_{53}}
=\left( \frac{\Delta\phi}{2\pi}\right)^2. \label{eq:BR1:conical45}
\end{eqnarray}
Moreover, the absence of conical singularities on $[z_5,\infty)$ requires the condition
\begin{eqnarray}
\lim_{\rho\to 0}\sqrt{\frac{\rho^2 f}{g_{\psi \psi } }}=\frac{\Delta\psi}{2\pi} 
\Longleftrightarrow C_f=\left( \frac{\Delta\psi}{2\pi}\right)^2, \label{eq:BR1:conical5}
\end{eqnarray}
and  from Eq.~(\ref{eq:BR1:conical5}), the choice of $C_f=1$  determines  $\Delta \psi = 2\pi$.
To avoid conical singularities on three $\phi$-axes, the periodicities given in Eqs.~(\ref{eq:BR1:conical01}), (\ref{eq:BR1:conical23}) and (\ref{eq:BR1:conical45}) must coincide with one another, which leads to
\begin{align}
\left( \frac{\Delta\phi}{2\pi}\right)^2=
\frac{2z_{10}^2z_{30}^2 z_{50} }
{z_{20}^2 z_{40}^2 }=
\frac{2z_{30}^2z_{32}^2z_{41}^2z_{50}z_{52}}
{z_{31}^2 z_{40}^2  z_{42}^2 z_{51} }=\frac{2 z_{50}z_{52}z_{54}  }
{z_{51}z_{53}}. \label{eq:BR1-conicalfree}
\end{align}

\medskip
To provide physical interpretations for these parameters, we introduce the positive parameters  $(\ell,r_1,r_2,\mu_1,\mu_2,\nu)$ as follows:
\begin{eqnarray}
z_0=- \ell,\ z_1=0,\ z_2=r_1 \ell,\ z_3=(r_1+\mu)\ell,\ z_4=(r_1+\mu+r_2)\ell,\ z_5=(r_1+\mu+r_2+\nu)\ell,
\end{eqnarray}
where $\ell$ represents the sizes of the expanding bubble on $[z_0,z_1]$,  $\mu\ell$ and $\nu\ell$ represent the sizes of the inner $\phi$-axis on $[z_2,z_3]$ between two black ring horizons  and the inner $\phi$-axis on $[z_4,z_5]$ inside the inner black ring, respectively. 
Additionally, $r_1 \ell$ and $r_2\ell$ denotes the sizes of the outer black ring horizon on $[z_1,z_2]$ and the inner black ring horizon on $[z_3,z_4]$, respectively.
In terms of these parameters, Eq.~(\ref{eq:BR1-conicalfree}) becomes
\begin{align}
&\frac{\mu ^2 \left(r_1+1\right){}^2 \left(\mu +r_1+r_2\right){}^2 \left(\mu +\nu +r_2\right)}{\left(\mu +r_1\right){}^2 \left(\mu +r_2\right){}^2
   \left(\mu +\nu +r_1+r_2\right)}=1, \label{eq:BR-conicalfree_a}\\
 &  \frac{\nu  \left(r_1+1\right){}^2 \left(\mu +r_1+r_2+1\right){}^2 \left(\mu +\nu +r_2\right)}{\left(\mu +r_1+1\right){}^2
   \left(\nu +r_2\right) \left(\mu +\nu +r_1+r_2\right)}=1,  \label{eq:BR-conicalfree_b}
\end{align}
\begin{align}
\left(\frac{\Delta\phi}{2\pi}\right)^2=\frac{2\ell(\mu+r_1+1)^2 (\mu+\nu+r_1+r_2+1)^2}{(r_1+1)^2 (\mu+r_1+r_2+1)^2},  \label{eq:BR-conicalfree_c}
\end{align}
where Eqs.~(\ref{eq:BR-conicalfree_a}) and (\ref{eq:BR-conicalfree_b}) relate the sizes of two black ring horizons to those of two inner $\phi$-axes, and   Eqs.~(\ref{eq:BR-conicalfree_c}) determines the periodicity of $\phi$.

\begin{figure}[h]
 \centering
\includegraphics[width=6.5cm]{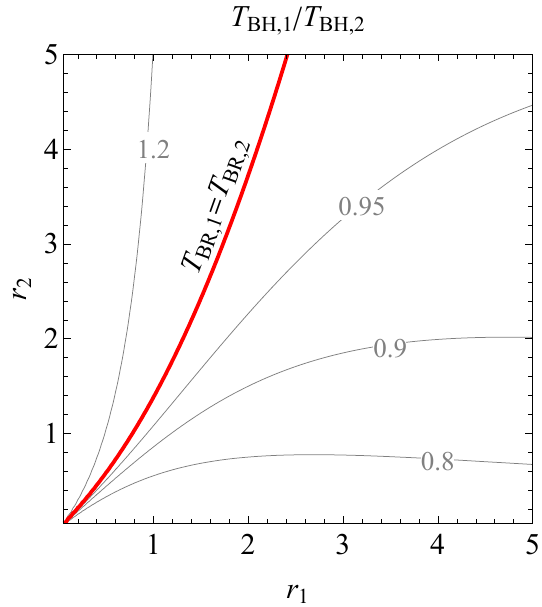}
\includegraphics[width=6.5cm]{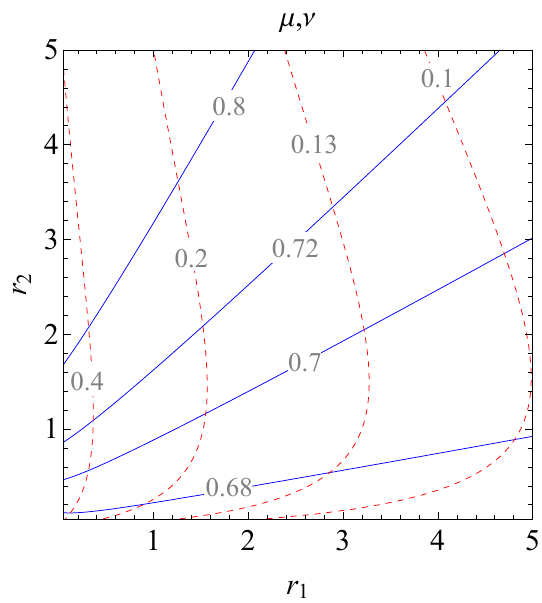}
\caption{
Phase diagram of the black di-ring in an expanding bubble.
The temperature ratio between the two black ring horizons and the values of $\mu$ and $\nu$ are given in the $(r_1, r_2)$ plane. In the right panel, the blue and red dashed curves correspond to constant curves of $\mu$ and $\nu$, respectively.
\label{fig:BR1_T}}
\end{figure}

\medskip
The temperatures $T_{\rm BR,1} (=:1/\beta_{\rm BR,1})$, $T_{\rm BR,2} (=:1/\beta_{\rm BR,2})$ and $T_{\rm AH} (=:1/\beta_{\rm AH})$ for the outer black ring horizon on $[z_1,z_2]$ and the inner  black ring horizon on  $[z_3,z_4]$ and the bubble acceleration horizon on $(-\infty,z_0]$ can be computed as, respectively, 
\begin{eqnarray}
&&\lim_{\rho\to 0}\sqrt{\frac{\rho^2 f}{-g_{tt } }}=\frac{\beta_{\rm BR,1}}{2\pi}  
\Longleftrightarrow 
\frac{z_{21}^2z_{30}^2z_{41}^2z_{50}}
{z_{20}^2z_{31}^2z_{40}^2z_{51}}=\left(\frac{\beta_{\rm BR,1}}{2\pi}\right)^2,\label{eq:BR1:conical12}\\
&&\lim_{\rho\to 0}\sqrt{\frac{\rho^2 f}{-g_{tt } }}=\frac{\beta_{\rm BR,2}}{2\pi}  
\Longleftrightarrow 
\frac{z_{41}^2z_{43}^2z_{50}z_{52}}{z_{40}^2z_{42}^2z_{51}z_{53}}=
\left(\frac{\beta_{\rm BR,2}}{2\pi}\right)^2,\label{eq:BR1:conical34}\\
&&\lim_{\rho\to 0}\sqrt{\frac{\rho^2 f}{-g_{tt } }}=\frac{\beta_{\rm AH}}{2\pi}  
\Longleftrightarrow 
1=\left(\frac{\beta_{\rm AH}}{2\pi}\right)^2.\label{eq:BR1:conical0}
\end{eqnarray}
In terms of the parameters $(r_1,r_2,\mu,\nu)$,  Eqs.~(\ref{eq:BR1:conical12}), (\ref{eq:BR1:conical34}) and (\ref{eq:BR1:conical0}) can be written as, respectively, 
\begin{eqnarray}
\left(\frac{\beta_{\rm BR,1}}{2\pi}\right)^2&=&\frac{r_1^2 (\mu+r_1+1)^2 (\mu+r_1+r_2)^2 (\mu+\nu+r_1+r_2+1)}{(r_1+1)^2 (\mu+r_1)^2 (\mu+r_1+r_2+1)^2 (+\nu+r_1+r_2)},\\
\left(\frac{\beta_{\rm BR,2}}{2\pi}\right)^2&=&  \frac{r_2^2 (\mu+\nu+r_2) (\mu+r_1+r_2)^2 (\mu+\nu+r_1+r_2+1)}{(\nu+r_2) (\mu+r_2)^2 (\mu+r_1+r_2+1)^2 (\mu+\nu+r_1+r_2)},\\
   \left(\frac{\beta_{\rm AH}}{2\pi}\right)^2&=&1.
\end{eqnarray}
We characterize the solution by the ratio of the  temperatures of the outer black ring to the inner black ring, which is given by
\begin{eqnarray}
\left(\frac{T_{\rm BR,1}}{T_{\rm BR,2}}\right)^2=\left(\frac{\beta_{\rm BR,2}}{\beta_{\rm BR,1}}\right)^2=\frac{r_2^2(r_1+1)^2  (\mu+r_1)^2 (\mu+\nu+r_2)}{r_1^2 (\nu+r_2) (\mu+r_1+1)^2 (\mu+r_2)^2}.
\end{eqnarray}
As shown in Fig.~\ref{fig:BR1_T}, the black di-ring can achieve static equilibrium in an expanding bubble for any  positive parameters $(r_1,r_2)$.
In either limit $r_1\to 0$ or $r_2\to 0$ where $T_{\rm BR,1}\to\infty$ or $T_{\rm BR,2}\to\infty$,  it is possible to reproduce a static black ring in an expanding bubble~\cite{Astorino:2022fge}.

\medskip
Despite the addition of another bubble acceleration horizon on the right end, represented as a semi-infinite timelike rod on $[z_6, \infty)$, the static black di-ring can still achieve balance.
However, having only the right acceleration horizon is insufficient to support the black di-ring due to unavoidable conical singularities.
This indicates that the gravitational attraction between the two black rings cannot counterbalance the pulling force along the $\psi$-rotational axis caused by the right accelerating horizon alone.
Therefore, the presence of the left acceleration horizon on $(-\infty,z_0]$ is necessary to achieve equilibrium.

\subsection{Bicycling black ring in expanding bubbles}\label{sec:obdr}
\begin{figure}[h]
 \centering
\includegraphics[width=9cm]{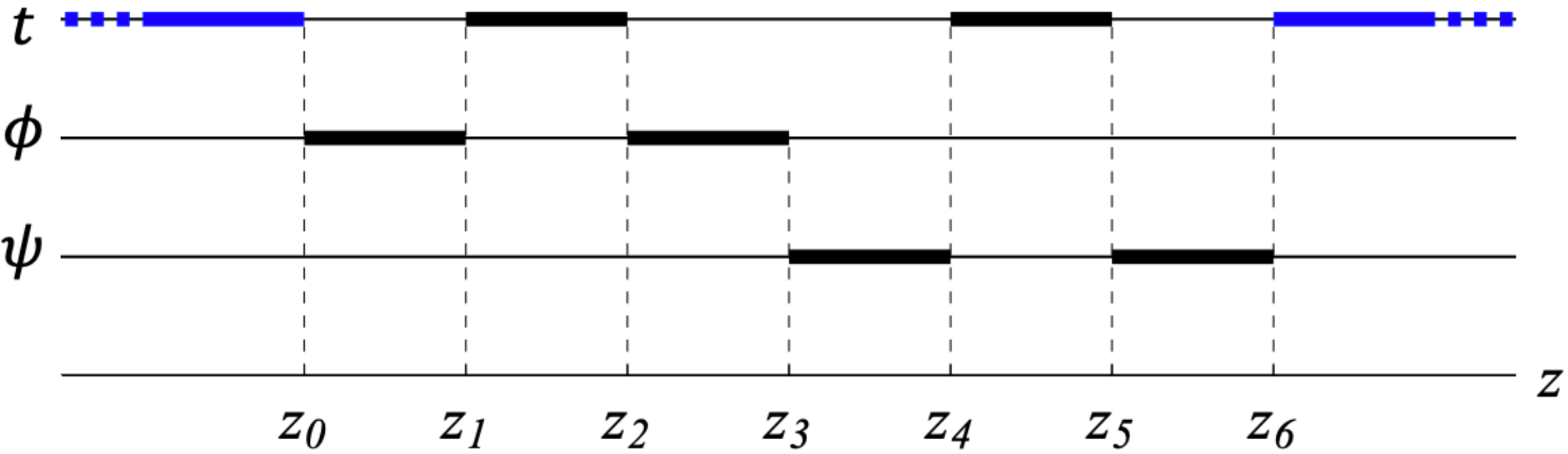}
\caption{Rod structure of a bicycling black ring inside expanding bubble. \label{fig:rod-OBDR}}
\end{figure}

Thirdly, we investigate ``a static bicycling black ring in an expanding bubble," as depicted in Fig.~\ref{fig:rod-OBDR}. 
In contrast to the previous two cases, where we assumed a single acceleration bubble horizon for the black saturn, we will observe that the bicycling black ring necessitates two bubble acceleration horizons on both sides for static equilibrium.
From Fig.~\ref{fig:rod-OBDR}, we can derive the metric as follows:
\begin{eqnarray}
ds^2=-\frac{\rho^2\mu_1\mu_4\mu_6}{\mu_0\mu_2\mu_5}dt^2
+\frac{\mu_0\mu_2}{\mu_1\mu_3}d\phi^2
+\frac{\mu_3\mu_5}{\mu_4\mu_6}d\psi^2+
f(d\rho^2+dz^2),
\end{eqnarray}
\begin{eqnarray}
f=C_f \frac{\mu_1\mu_4\mu_6}{\mu_0\mu_2\mu_5}
\frac{R_{01}^2  R_{03} R_{04} R_{06}     R_{12}^2  R_{15}     R_{23}  R_{24} R_{26}    R_{34}R_{36}   R_{45}^2  R_{56}^2}
{R_{00}R_{02}^2R_{05}  R_{11}R_{13}R_{14}R_{16}  R_{22}R_{25}  R_{33}R_{35}  R_{44}R_{46}^2  R_{55}R_{66} }.
\end{eqnarray}
This spacetime has two black ring horizons with topology $S^1\times S^2$ on the timelike finite rods $[z_1,z_2]$ and $[z_4,z_5]$, where they are concentric and in the orthogonal planes. Additionally, there exist two bubble acceleration horizons with topology $S^1\times \mathbb{R}^2$ on the timelike semi-infinite rods $(-\infty,z_0]$ and $[z_6,\infty)$. Furthermore, there are two expanding bubbles with topology $S^1\times \mathbb{R}$ on the spacelike finite rods $[z_0,z_1]$ and $[z_5,z_6]$. The two orthogonal inner $\phi$-rotational and $\psi$-rotational axes on the spacelike finite rods $[z_2,z_3]$ and $[z_3,z_4]$ have topology $D^2$, tangential at the point $z=z_3$.

In order to remove conical singularities on four rods $[z_0,z_1]$, $[z_2,z_3]$, $[z_3,z_4]$ and $[z_5,z_6]$ on the $z$-axis, we impose the regularity conditions as, respectively,  
\begin{eqnarray}
&&\lim_{\rho\to 0}\sqrt{\frac{\rho^2 f}{g_{\phi \phi } }}=\frac{\Delta\phi}{2\pi} \Longleftrightarrow 
4C_f
\frac{z_{10}^2z_{30} z_{40} z_{60}}
{z_{20}^2 z_{50} }
=\left( \frac{\Delta\phi}{2\pi}\right)^2,
\label{eq:OBD1:conical01}\\
&&\lim_{\rho\to 0}\sqrt{\frac{\rho^2 f}{g_{\phi \phi } }}=\frac{\Delta\phi}{2\pi} \Longleftrightarrow 
4C_f
\frac{z_{30}z_{32} z_{40} z_{42}  z_{51}  z_{60}z_{62} }
{z_{31} z_{41}   z_{50}z_{52}  z_{61}}
=\left( \frac{\Delta\phi}{2\pi}\right)^2,
\label{eq:OBD1:conical23}
\end{eqnarray}
and 
\begin{eqnarray}
&&\lim_{\rho\to 0}\sqrt{\frac{\rho^2 f}{g_{\psi \psi } }}=\frac{\Delta\psi}{2\pi} \Longleftrightarrow 
4C_f\frac{ z_{40}z_{42}z_{43}   z_{51}  z_{60}z_{62}z_{63}   }
{z_{41}   z_{50}z_{52}z_{53}z_{61} }=\left( \frac{\Delta\psi}{2\pi}\right)^2,
\label{eq:OBD1:conical34}\\
&&\lim_{\rho\to 0}\sqrt{\frac{\rho^2 f}{g_{\psi \psi } }}=\frac{\Delta\psi}{2\pi} \Longleftrightarrow 
4C_f\frac{z_{60}z_{62}z_{63}z_{65}^2   }
{z_{61}z_{64}^2 }=\left( \frac{\Delta\psi}{2\pi}\right)^2.
\label{eq:OBD1:conical56}
\end{eqnarray}
From Eqs.~(\ref{eq:OBD1:conical01}), (\ref{eq:OBD1:conical23}) and Eqs.~(\ref{eq:OBD1:conical34}), (\ref{eq:OBD1:conical56}), we have
\begin{align}
\left( \frac{\Delta\phi}{2\pi}\right)^2=
\frac{z_{10}^2z_{30} z_{40} z_{60}}
{z_{20}^2 z_{50} }
=
\frac{z_{30}z_{32} z_{40} z_{42}  z_{51}  z_{60}z_{62} }
{z_{31} z_{41}   z_{50}z_{52}  z_{61}},\label{eq:OBD1:conical-phi}
\end{align}
and 
\begin{align}
\left( \frac{\Delta\psi}{2\pi}\right)^2=\frac{z_{60}z_{62}z_{63}z_{65}^2   }
{z_{61}z_{64}^2 }=\frac{ z_{40}z_{42}z_{43}   z_{51}  z_{60}z_{62}z_{63}   }
{z_{41}   z_{50}z_{52}z_{53}z_{61} },\label{eq:OBD1:conical-psi}
\end{align}
where we have chosen $C_f=1/4$.

\medskip
The temperatures $T_{\rm BR,1} (=:1/\beta_{\rm BR,1})$, $T_{\rm BR,2} (=:1/\beta_{\rm BR,2})$,  $T_{\rm AH,1} (=:1/\beta_{\rm AH,1})$ and $T_{\rm AH,2} (=:1/\beta_{\rm AH,2})$ for the left black ring horizon on $[z_1,z_2]$ and the right black ring horizon on  $[z_4,z_5]$ and the bubble acceleration horizons on $(-\infty,z_0]$  and $[z_6,\infty)$  can be written as, respectively, 
\begin{eqnarray}  
&&\lim_{\rho\to 0}\sqrt{\frac{\rho^2 f}{-g_{tt } }}=\frac{\beta_{\rm BR,1}}{2\pi}  
\Longleftrightarrow 
\frac{z_{21}^2z_{30}z_{40}z_{51}z_{60}}
{4z_{20}^2z_{31}z_{41}z_{50}z_{61}}=\left(\frac{\beta_{\rm BR,1}}{2\pi}\right)^2,
\label{eq:OBD1:conical12}\\
&&\lim_{\rho\to 0}\sqrt{\frac{\rho^2 f}{-g_{tt} }}=\frac{\beta_{\rm BR, 2}}{2\pi} 
\Longleftrightarrow 
\frac{ z_{51}z_{54}^2 z_{60}z_{62}z_{63} }
{4z_{50}z_{52}z_{53}z_{61}z_{64}^2}=\left( \frac{\beta_{\rm BR, 2}}{2\pi}\right)^2,
\label{eq:OBD1:conical45}\\
&& \lim_{\rho\to 0}\sqrt{\frac{\rho^2 f}{-g_{tt } }}=\frac{\beta_{\rm AH,1}}{2\pi} \Longleftrightarrow 
\frac{1}{4}=\left(\frac{\beta_{\rm AH,1}}{2\pi}\right)^2,\\
&& \lim_{\rho\to 0}\sqrt{\frac{\rho^2 f}{-g_{tt } }}=\frac{\beta_{\rm AH,2}}{2\pi} \Longleftrightarrow \frac{1}{4}=\left( \frac{\beta_{\rm AH,2}}{2\pi}\right)^2.
\end{eqnarray}

\begin{figure}[h]
\centering
\begin{minipage}{0.45\columnwidth}
\includegraphics[width=6.3cm]{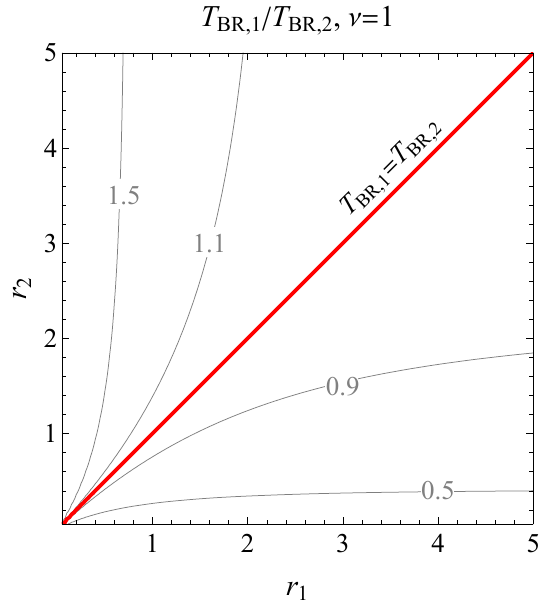}
\includegraphics[width=6.3cm]{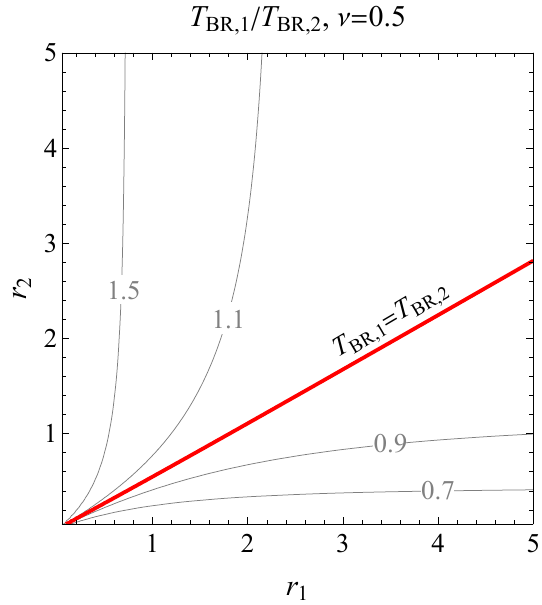}
\end{minipage}
\begin{minipage}{0.45\columnwidth}
\includegraphics[width=6.3cm]{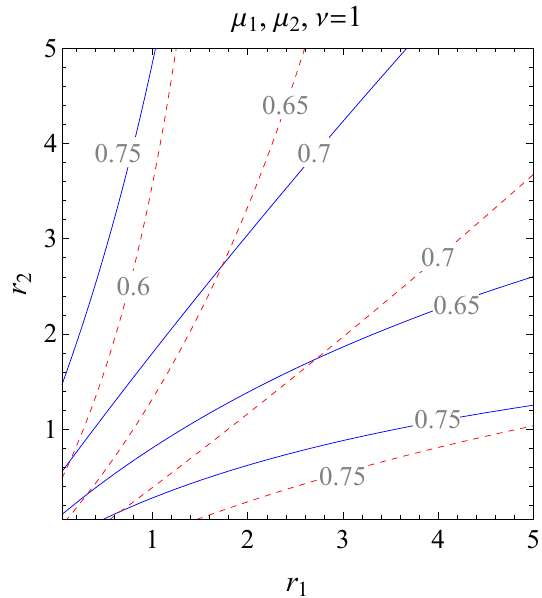}
\includegraphics[width=6.3cm]{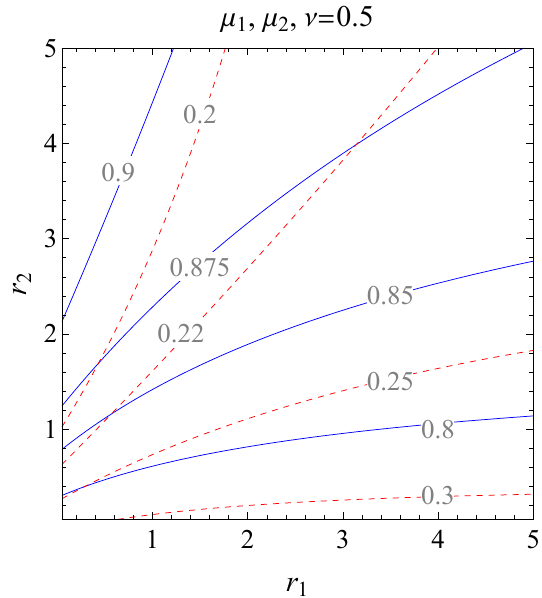}
\end{minipage}
\caption{The phase diagram of the bicycling black ring in expanding bubbles for $\nu=1$ and $\nu=0.5$ illustrated in the ($r_1,r_2$) plane. The temperature ratio between two ring horizons and the values of $\mu_1$ (blue curves) and $\mu_2$ (red dashed curve) are shown in the $(r_1,r_2)$ plane. \label{fig:ODR2}}
\end{figure}

\medskip
Instead of the rod endpoints $z_i\ (i=0,\ldots,6)$, we use the following positive parameters $(\ell,r_1,r_2,\mu_1,\mu_2,\nu)$:
\begin{eqnarray}
\begin{split}
&z_0=- (1+r_1+\mu_1) \ell,\quad z_1=-(r_1+\mu_1) \ell,\quad z_2=-\mu_1 \ell,\quad z_3=0, \\
& z_4=\mu_2 \ell,\quad z_5=(r_2+\mu_2)\ell,\quad
z_6 = (r_2+\mu_2+\nu)\ell,
\end{split}
\end{eqnarray}
where $\ell$ and $\ell \nu$ represent the sizes of the expanding bubbles on $[z_0,z_1]$ and $[z_5,z_6]$, respectively,  $\mu_1\ell$ and $\mu_2\ell$ represent the sizes of the two inner $\phi$ and $\psi$ axes on $[z_2,z_3]$ and $[z_3,z_4]$, respectively. 
Additionally, $r_1 \ell$ and $r_2\ell$ denotes the sizes of the left black ring horizon on $[z_1,z_2]$ and right black ring horizon on $[z_4,z_5]$, respectively.
From Eqs.~(\ref{eq:OBD1:conical-phi}) and (\ref{eq:OBD1:conical-psi}), one can obtain
the balance conditions
\begin{align}
&\frac{\mu _1 \left(\mu _1+\mu _2\right) \left(r_1+1\right){}^2 \left(\mu _1+\mu
   _2+r_1+r_2\right) \left(\mu _1+\mu _2+\nu +r_2\right)}{\left(\mu _1+r_1\right)
   \left(\mu _1+\mu _2+r_1\right) \left(\mu _1+\mu _2+r_2\right) \left(\mu _1+\mu
   _2+\nu +r_1+r_2\right)}=1,\label{eq:OBD1:conical-phi-1}\\
   &\frac{\mu _2 \left(\mu _1+\mu _2\right) \left(\mu _1+\mu _2+r_1+1\right) \left(\mu
   _1+\mu _2+r_1+r_2\right) \left(\nu +r_2\right){}^2}{\nu ^2 \left(\mu _2+r_2\right)
   \left(\mu _1+\mu _2+r_1\right) \left(\mu _1+\mu _2+r_2\right) \left(\mu _1+\mu
   _2+r_1+r_2+1\right)}=1.\label{eq:OBD1:conical-psi-1}
\end{align}
To find a solution, we first consider the simple case where the spacetime becomes symmetric under the exchange $\phi \leftrightarrow  \psi$
\begin{align}
r:=r_1=r_2,\quad \mu := \mu_1 = \mu_2 ,\quad \nu = 1,
\end{align}
with which Eqs.~(\ref{eq:OBD1:conical-phi-1}) and (\ref{eq:OBD1:conical-psi-1}) reduce to a single condition
\begin{align}
\frac{ 4 \mu^2(1+r)^2  (1+r+2\mu)}{(r+2\mu)^2(1+2r+2\mu)}=1.
\end{align}
This equation admits a simple solution:
\begin{align}
r = \frac{8 \mu ^3+12 \mu ^2-10 \mu
   -1+(2 \mu -1)^{3/2} \sqrt{8 \mu ^3+4 \mu ^2+6 \mu -1}}{4 \left(1-2 \mu ^2\right)},\quad \frac{1+\sqrt{17}}{8}<\mu<\frac{1}{\sqrt{2}},
\end{align}
where $r$ takes the value between $0$ and $\infty$ for the possible range of $\mu$.
In more general cases, one can find the solution for any set of $(\nu,r_1,r_2)$ by solving Eqs.~(\ref{eq:OBD1:conical-phi-1}) and~(\ref{eq:OBD1:conical-psi-1}) numerically.
In Fig.~\ref{fig:ODR2}, we illustrate the phase diagram for $\nu=0.5$ and $\nu=1$.

Taking the limit $z_{6}\to \infty$ with the rescaling $\phi \to z_6^{1/2} \phi $ and $\psi \to z_6 \psi$, the right bubble acceleration horizon is eliminated, and Eq.~(\ref{eq:OBD1:conical-psi}) simplifies to
\begin{align}
\frac{ z_{40}z_{42}z_{43} z_{51} }
{z_{41} z_{50}z_{52}z_{53}}=1.
\end{align}
We can verify that the left-hand side is always less than unity, indicating an unavoidable conical singularity on $[z_3,z_4]$.
Since the system is symmetric under $\psi \leftrightarrow \phi$ and $z_{i} \leftrightarrow z_{6-i}$, the same holds for the limit $z_0\to-\infty$.
Therefore, the bicycling black ring cannot be supported by a single acceleration horizon alone.

\subsection{Five-dimensional black hole binary in expanding bubbles}\label{sec:5dbbh}
\begin{figure}[h]
 \centering
\includegraphics[width=9cm]{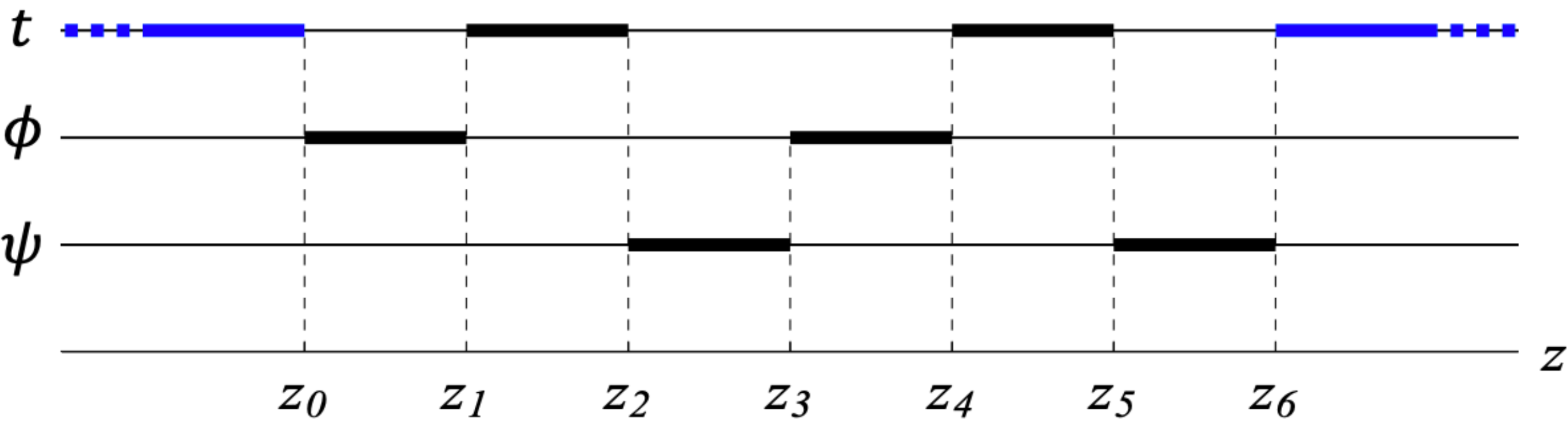}
\caption{Rod structure of a five-dimensional black hole binary in expanding bubbles. \label{fig:BBH-rod}}
\end{figure}

Finally, we examine ``a static five-dimensional black hole binary in expanding bubbles," as illustrated in Fig.~\ref{fig:BBH-rod}. Similar to the previous bicycling black ring, which requires the presence of two acceleration bubble horizons to achieve static equilibrium, we will find that a five-dimensional black hole binary also requires two bubble acceleration horizons on both sides for equilibrium. 
From Fig.~\ref{fig:BBH-rod}, we can derive the metric as follows:
\begin{eqnarray}
ds^2=-\frac{\rho^2\mu_1\mu_4\mu_6}{\mu_0\mu_2\mu_5}dt^2
+\frac{\mu_0\mu_3}{\mu_1\mu_4}d\phi^2
+\frac{\mu_2\mu_5}{\mu_3\mu_6}d\psi^2+
f(d\rho^2+dz^2),
\end{eqnarray}
with
\begin{eqnarray}
f=C_f \frac{\mu_1\mu_4\mu_6}{\mu_0\mu_2\mu_5}
\frac{R_{01}^2  R_{04}^2 R_{06}     R_{12}R_{13}R_{15}     R_{23}R_{24} R_{26}^2    R_{34}R_{35}   R_{45}  R_{56}^2}
{R_{00}R_{02}R_{03}R_{05}  R_{11}R_{14}^2R_{16}  R_{22}R_{25}^2  R_{33}R_{36}  R_{44}R_{46}  R_{55}R_{66} }.
\end{eqnarray}

This spacetime has two black hole horizons with topology $S^3$ located on the timelike finite rods $[z_1,z_2]$ and $[z_4,z_5]$.
Additionally, there are two bubble acceleration horizons with topology  $S^1\times \mathbb{R}^2$ positioned on the timelike semi-infinite rods $(-\infty,z_0]$ and $[z_6,\infty)$.
Furthermore, there are two expanding bubbles with topology $S^1\times \mathbb{R}$ situated on the spacelike finite rods $[z_0,z_1]$ and $[z_5,z_6]$.
The two orthogonal inner $\psi$-rotational and $\phi$-rotational axes have  topology $D^2$ located on $[z_2,z_3]$ and $[z_3,z_4]$, orthogonal and tangential at the point $z=z_3$.

\medskip
In order to remove conical singularities on four rods $[z_0,z_1]$, $[z_2,z_3]$, $[z_3,z_4]$ and $[z_5,z_6]$ on the $z$-axis, we impose the regularity conditions as, respectively,  
\begin{eqnarray}
&&\lim_{\rho\to 0}\sqrt{\frac{\rho^2 f}{g_{\phi \phi } }}=\frac{\Delta\phi}{2\pi} \Longleftrightarrow 
4C_f
\frac{z_{10}^2 z_{40}^2 z_{60}}
{z_{20} z_{30} z_{50} }
=\left( \frac{\Delta\phi}{2\pi}\right)^2,
\label{eq:BBH-period-01}\\
&&\lim_{\rho\to 0}\sqrt{\frac{\rho^2 f}{g_{\psi \psi } }}=\frac{\Delta\psi}{2\pi} \Longleftrightarrow 
4C_f
\frac{z_{31}z_{32} z_{40}^2 z_{42}  z_{51}  z_{60}z_{62}^2 }
{z_{30} z_{41}^2   z_{50}z_{52}^2  z_{61}}
=\left( \frac{\Delta\psi}{2\pi}\right)^2,
\label{eq:BBH-period-23}\\
&&\lim_{\rho\to 0}\sqrt{\frac{\rho^2 f}{g_{\phi \phi } }}=\frac{\Delta\phi}{2\pi} \Longleftrightarrow 
4C_f\frac{ z_{40}^2z_{42}z_{43}   z_{51}z_{53}  z_{60}z_{62}^2   }
{z_{41}^2   z_{50}z_{52}^2 z_{61}z_{63} }=\left( \frac{\Delta\phi}{2\pi}\right)^2,
\label{eq:BBH-period-34}\\
&&\lim_{\rho\to 0}\sqrt{\frac{\rho^2 f}{g_{\psi \psi } }}=\frac{\Delta\psi}{2\pi} \Longleftrightarrow 
4C_f\frac{z_{60}z_{62}^2z_{65}^2   }
{z_{61}z_{63}z_{64} }=\left( \frac{\Delta\psi}{2\pi}\right)^2\label{eq:BBH-period-56}.
\end{eqnarray}
By choosing $C_f=\frac{1}{4}$, from Eqs.~(\ref{eq:BBH-period-01}), (\ref{eq:BBH-period-34}) and Eqs.~(\ref{eq:BBH-period-23}), (\ref{eq:BBH-period-56}), we have, respectively, 
\begin{align}
\left( \frac{\Delta\phi}{2\pi}\right)^2=
\frac{z_{10}^2 z_{40}^2 z_{60}}
{z_{20} z_{30} z_{50} }
=\frac{ z_{40}^2z_{42}z_{43}   z_{51}z_{53}  z_{60}z_{62}^2   }
{z_{41}^2   z_{50}z_{52}^2 z_{61}z_{63} },\label{eq:BBH-cf-phi}
\end{align}
and
\begin{align}
\left( \frac{\Delta\psi}{2\pi}\right)^2=
\frac{z_{31}z_{32} z_{40}^2 z_{42}  z_{51}  z_{60}z_{62}^2 }
{z_{30} z_{41}^2   z_{50}z_{52}^2  z_{61}}
=\frac{z_{60}z_{62}^2z_{65}^2   }
{z_{61}z_{63}z_{64} }.\label{eq:BBH-cf-psi}
\end{align}

\medskip
The temperatures $T_{\rm BH,1} (=:1/\beta_{\rm BH,1})$, $T_{\rm BH,2} (=:1/\beta_{\rm BH,2})$, $T_{\rm AH,1} (=:1/\beta_{\rm AH,1})$ and $T_{\rm AH,2} (=:1/\beta_{\rm AH,2})$ for the left black hole horizon on $[z_1,z_2]$ and the right black hole horizon on  $[z_4,z_5]$ and the bubble acceleration horizons on $(-\infty,z_0]$  and $[z_6,\infty)$  can be written as, respectively, 
\begin{eqnarray}
&&\lim_{\rho\to 0}\sqrt{\frac{\rho^2 f}{-g_{tt } }}=\frac{\beta_{BH,1}}{2\pi} \Longleftrightarrow 
\frac{z_{21}z_{31}z_{40}^2z_{51}z_{60}}
{4z_{20} z_{30}z_{41}^2 z_{50}z_{61} }
=\left( \frac{\beta_{BH,1}}{2\pi}\right)^2,
\label{eq:BBH-period-12}\\
&&\lim_{\rho\to 0}\sqrt{\frac{\rho^2 f}{-g_{tt } }}=\frac{\beta_{BH,2}}{2\pi} \Longleftrightarrow 
\frac{z_{51}z_{53}z_{62}^2z_{54}z_{60}}
{4z_{50} z_{61}z_{52}^2 z_{63}z_{64} }
=\left( \frac{\beta_{BH,2}}{2\pi}\right)^2,
\label{eq:BBH-period-45}\\
&&\lim_{\rho\to 0}\sqrt{\frac{\rho^2 f}{-g_{tt} }}=\frac{\beta_{AH,1}}{2\pi} \Longleftrightarrow 
\frac{1}{4}= \left( \frac{\beta_{AH,1}}{2\pi}\right)^2,
\label{eq:BBH-period-0}\\
&&\lim_{\rho\to 0}\sqrt{\frac{\rho^2 f}{-g_{tt } }}=\frac{\beta_{AH,2}}{2\pi} \Longleftrightarrow 
\frac{1}{4}= \left( \frac{\beta_{AH,2}}{2\pi}\right)^2.
\label{eq:BBH-period-6}
\end{eqnarray}

\begin{figure}[h]
 \centering
  \begin{minipage}{0.45\columnwidth}
\includegraphics[width=6.3cm]{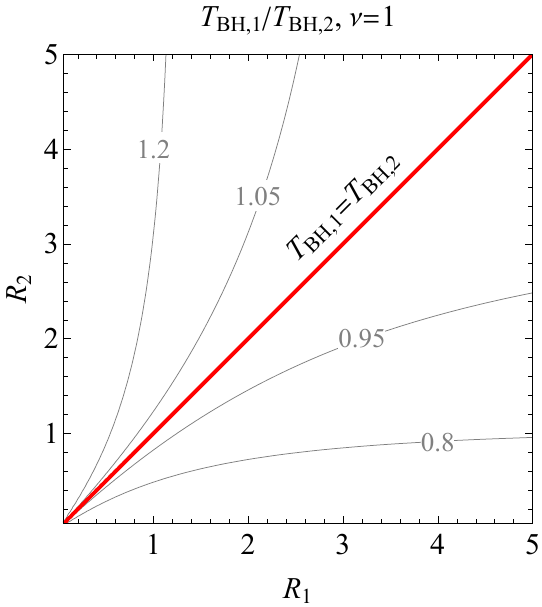}
\includegraphics[width=6.3cm]{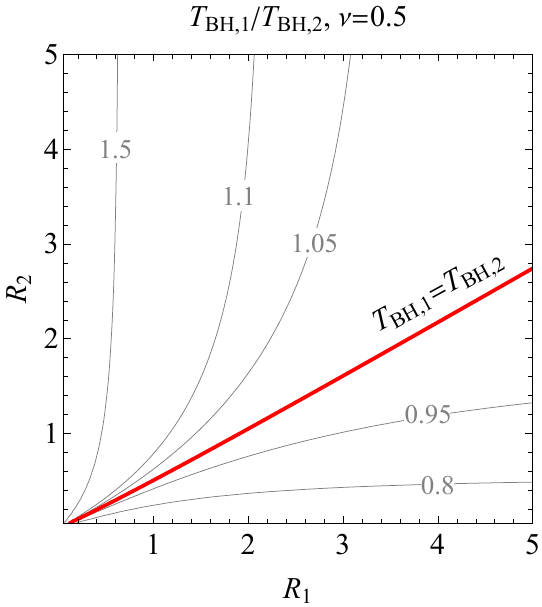}
\end{minipage}
 \begin{minipage}{0.45\columnwidth}
\includegraphics[width=6.3cm]{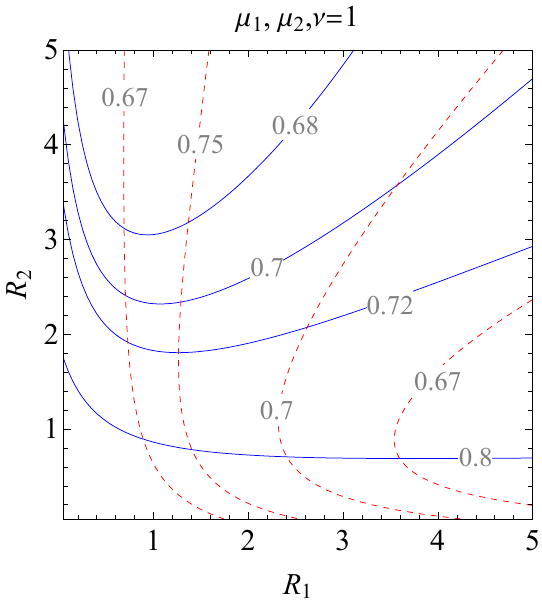}\hspace{5mm}
\includegraphics[width=6.3cm]{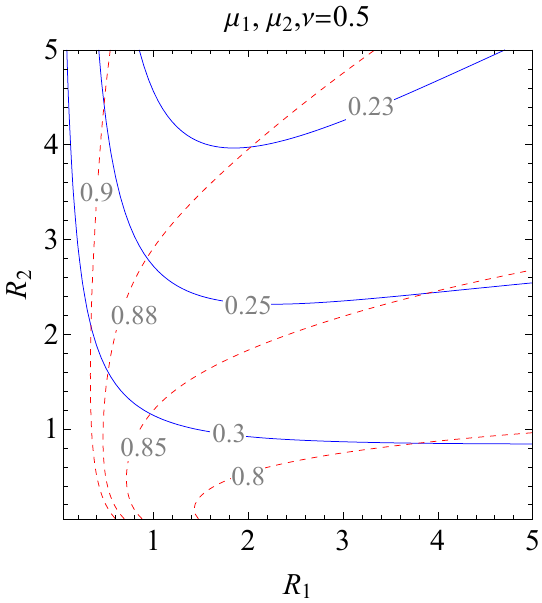}
\end{minipage}
\caption{The ratio of the black hole temperature between two black holes in $(\mu_1,\mu_2)$ plane for $\gamma=0.5,1$. The parameters $\mu_1$ (blue curves) and $\mu_2$ (red dashed curves) for given $(R_1,R_2)$ and $\nu=0.5,1$. \label{fig:bbh-nu1nu2}}
\end{figure}

To solve the regularity conditions~(\ref{eq:BBH-cf-phi}) and (\ref{eq:BBH-cf-psi}), we introduce the positive parameters $(\ell,R_1,R_2,\mu_1,\mu_2,\nu)$ as
\begin{align}
\begin{split}
 &z_0 = -(1+R_1+\mu_1)\ell,\quad  z_1 = -(R_1+\mu_1),\quad z_2 = -\mu_1 \ell,\quad z_3 =0,\\
&  z_4 = \mu_2 \ell,\quad z_5 = (R_2+\mu_2)\ell,\quad z_6 = (\nu+R_2+\mu_2)\ell,
\end{split}
\end{align}
where $\ell$ and $\ell \nu$ represent the sizes of the expanding bubbles on $[z_0,z_1]$ and $[z_5,z_6]$, respectively,  $\mu_1\ell$ and $\mu_2\ell$ represent the sizes of the two inner $\psi$ and $\phi$ axes on $[z_2,z_3]$ and $[z_3,z_4]$, respectively. 
Additionally, $R_1 \ell$ and $R_2\ell$ denotes the sizes of the left black hole horizon on $[z_1,z_2]$ and the right black hole horizon on $[z_4,z_5]$, respectively.
In terms of these parameters, the conditions~(\ref{eq:BBH-cf-phi}) and (\ref{eq:BBH-cf-psi}) can be rewritten as
\begin{align}
&\frac{\mu _2 \left(\mu _1+\mu _2\right) \left(R_1+1\right) \left(\mu _1+R_1+1\right) \left(\mu _2+R_2\right) \left(\mu _1+\mu _2+R_1+R_2\right) \left(\mu _1+\mu
   _2+\nu +R_2\right){}^2}{\left(\mu _1+\mu _2+R_1\right){}^2 \left(\mu _1+\mu _2+R_2\right){}^2 \left(\mu _2+\nu +R_2\right) \left(\mu _1+\mu _2+\nu
   +R_1+R_2\right)}=1,\label{eq:BBH-con1}\\
   &\frac{\mu _1 \left(\mu _1+\mu _2\right) \left(\mu _1+R_1\right) \left(\mu _1+\mu _2+R_1+1\right){}^2 \left(\mu _1+\mu _2+R_1+R_2\right) \left(\nu +R_2\right)
   \left(\mu _2+\nu +R_2\right)}{\nu ^2 \left(\mu _1+R_1+1\right) \left(\mu _1+\mu _2+R_1\right){}^2 \left(\mu _1+\mu _2+R_2\right){}^2 \left(\mu _1+\mu
   _2+R_1+R_2+1\right)}=1,\label{eq:BBH-con2}
\end{align}
and the temperature ratio between two black holes can be expressed as
\begin{align}
\left(\frac{T_{BH,1}}{T_{BH,2}}\right)^2=\left(\frac{\beta_{BH,2}}{\beta_{BH,1}}\right)^2 =\frac{\left(R_1+1\right) R_2 \left(\mu _1+R_1+1\right) \left(\mu _2+R_2\right) \left(\mu _1+\mu _2+R_1\right){}^2 \left(\mu _1+\mu _2+\nu +R_2\right){}^2}{R_1
   \left(\mu _1+R_1\right) \left(\mu _1+\mu _2+R_1+1\right){}^2 \left(\mu _1+\mu _2+R_2\right){}^2 \left(\nu +R_2\right) \left(\mu _2+\nu +R_2\right)}.
\end{align}
By solving Eqs.~(\ref{eq:BBH-con1}) and (\ref{eq:BBH-con2}) numerically, we obtain a unique solution for any set of $(\nu,R_1,R_2)$ as shown in Fig.~\ref{fig:bbh-nu1nu2}.

\medskip
One can see that the presence of the  acceleration bubble horizons on both sides are inevitable to balance two black holes.  
To see this, one can eliminate one bubble acceleration horizon of two by taking the limit  $z_0\to -\infty$ or $z_6 \to \infty$ together with appropriate scalings of $\phi$ and $\psi$. 
Utilizing the symmetry under the exchange $\psi \leftrightarrow \phi$ and $z_i \leftrightarrow z_{6-i}$, we focus exclusively on the limit where $z_6 \to \infty$. 
In this limit, the condition (\ref{eq:BBH-cf-psi}) can be written as
\begin{align}
\frac{z_{31}z_{32}z_{40}^2 z_{42}z_{51}}{z_{30}z_{41}^2z_{50}z_{52}^2}=1.
\end{align}
However, it becomes evident that the numerator on the left-hand side is smaller than the denominator. 
This observation leads to the conclusion that the existence of conical singularities cannot be avoided when dealing with a single acceleration bubble horizon.

\section{Summary}\label{sec:summary}
In this paper, we have explored possible configurations for vacuum multi-horizon black holes that maintain static equilibrium within expanding bubbles.
 When the spacetime is described by the Weyl metric, solutions are readily derived from the provided rod structure.
We have specifically considered spacetimes with one or two acceleration horizons attached to either one or both ends of the rod structure. 
We have found that two bubbles, a black saturn, a black di-ring, a bicycling black ring, and a five-dimensional black hole binary can achieve static equilibrium within expanding bubbles. 
As discussed in Ref.~\cite{Astorino:2022fge}, the attraction from the acceleration horizon behaves like the de Sitter expansion force.
For the black saturn and the  black di-ring, the presence of only the right acceleration horizon is insufficient to support the two horizons due to the presence of unavoidable conical singularities. 
This indicates that the gravitational attraction between the black hole and black ring cannot counterbalance the pulling force along the $\psi$-rotational axis caused by the accelerating horizon alone. Therefore, achieving static equilibrium necessitates the presence of the left acceleration horizon.
Meanwhile, 
for the bicycling black ring and the five-dimensional black hole binary, having only a single acceleration horizon, whether it is  right or left,  is inadequate for achieving static equilibrium, resulting in the unavoidable appearance of conical singularities. 
This means that the gravitational attraction between the two horizons is insufficient to counterbalance the pulling force along either the $\psi$-rotational axis or the $\phi$-rotational axis caused by the single acceleration horizon. 
Therefore, the presence of the second acceleration horizon is necessary to achieve static equilibrium.

\medskip
In this paper, for simplicity, we have dealt with only the case of multi-black holes described by a diagonal metric. 
We can also consider possible generalization of these solutions to a rotational case and a charged case. 
Moreover, our results suggest that in five-dimensional Einstein equations with a positive cosmological constant, there exist  exact solutions describing static multi-black hole configuration such as a black saturn, a  black di-ring, a bicycling black ring and a five-dimensional black hole binary, which can realize static equilibrium by balancing gravitational force and  cosmological expansion. We leave these interesting issues for our future work.




\acknowledgments
We thank Takashi Mishima for useful comments.
RS was supported by JSPS KAKENHI Grant Number JP18K13541. 
ST was supported by JSPS KAKENHI Grant Number 21K03560.




\end{document}